\documentclass[a4paper,aps,pre,superscriptaddress,floatfix,nofootinbib,longbibliography,notitlepage,onecolumn]{revtex4-1}

\usepackage[utf8]{inputenc}
\usepackage[english]{babel}

\usepackage{amsmath,amssymb,amsfonts,amsthm,mathrsfs,amsopn}
\usepackage{mathtools}
\usepackage[percent]{overpic}
\usepackage{graphicx}
\usepackage{xcolor}
\usepackage{bm}
\usepackage{braket}
\usepackage{pifont}
\usepackage{blkarray,bigstrut}
\usepackage{multirow}
\usepackage{array}
\usepackage{booktabs}
\usepackage{dcolumn}
\usepackage{float}
\usepackage{svg}
\usepackage{comment}
\usepackage{verbatim}

\usepackage{hyperref}

\makeatletter
\renewcommand{\[}{\begin{equation}}
\renewcommand{\]}{\end{equation}}
\makeatother

\def\be{\begin{equation}}
\def\ee{\end{equation}}
\def\bc{\begin{center}}
\def\ec{\end{center}}
\def\bea{\begin{eqnarray}}
\def\eea{\end{eqnarray}}

\theoremstyle{plain}

\definecolor{brickred}{rgb}{0.7, 0.25, 0.33}
\definecolor{applegreen}{rgb}{0.55, 0.71, 0.0}

\begin{document}

\title{Swarming and Opinion Dynamics}
\author{Rommel Tchinda Djeudjo}
\affiliation{Department of Mathematics \& naXys, Namur Institute for Complex Systems, University of Namur, Rue Grafé 2, B5000 Namur, Belgium}
\author{Dibakar Ghosh}
\affiliation{Physics and Applied Mathematics Unit, Indian Statistical Institute, Kolkata 700108, India}
\author{Timoteo Carletti}
\affiliation{Department of Mathematics \& naXys, Namur Institute for Complex Systems, University of Namur, Rue Grafé 2, B5000 Namur, Belgium}

\begin{abstract}
Collective dynamics in multi-agent systems provide a powerful framework for understanding how coherent group-level patterns can emerge from simple interactions between individuals. Such phenomena are observed in many natural and artificial systems, including animal groups, robotic swarms, and distributed decision-making processes. In many situations, agents are not only characterized by their spatial motion, but also by internal states, e.g., opinions or preferences, which evolve through interactions with peers. Understanding how these internal states influence collective motion, and how spatial organization in turn affects internal dynamics, remains an important challenge. In this work, we propose a model of coupled collective motion and opinion dynamics. The spatial dynamics are governed by attraction--repulsion interactions, while the internal dynamics are described by a Deffuant-type opinion model. Our results show that the confidence threshold of the opinion dynamics plays a key role in controlling the number of opinion clusters, whereas the strength of the opinion-dependent spatial attraction determines whether these clusters spatially merge or remain separated. In addition, for the full-consensus state, we derive the expression for the radius of the stationary swarm distribution when a nonlinear attraction kernel is used, using a semi-analytical approach. The proposed framework may be useful for studying collective decision-making, animal group behavior, and coordination strategies in swarm robotics.
\end{abstract}

\maketitle
\section{Introduction}
\label{sec:intro}

Collective motion is one of the most remarkable phenomena observed in natural and artificial multi-agent systems. In nature, many living organisms are able to organize their motion in groups: birds form flocks, fish swim in schools, sheep and deer move in herds, while insects and microorganisms may produce complex collective patterns \citep{Bialek2012,Hemelrijk2012,Sumpter2010,Okubo1986}. These phenomena are particularly interesting because the global organization of the group emerges without any centralized control. Instead, each individual interacts mainly with its local neighbors, and these simple local interaction rules can generate coherent motion at the collective level. From a biological point of view, moving in groups provides several advantages. It may facilitate migration, increase the ability to find food, improve communication between individuals, and reduce the risk of predation \citep{Sumpter2010,Norris1988}. However, beyond the question of why such collective behaviors occur, a central scientific challenge is to understand how do they emerge. This question has attracted researchers from many disciplines, including biology, physics, mathematics, computer science, and control theory \citep{Toner1998,Toner2005,Cucker2007,Vicsek2012}. The study of collective motion is also important for technological applications, especially in swarm robotics, distributed decision-making, communication networks, and artificial multi-agent systems \citep{Miller2010,VanDerHoek2008,Brambilla2013,Kennedy2006}.

One of the earliest influential model of collective motion was introduced by Reynolds in 1987 through the concept of \emph{boids} \citep{Reynolds1987}. In this model, each artificial agent follows three basic behavioral rules: separation, which prevents collisions with nearby agents; alignment, which makes an agent adjust its direction according to its neighbors; and cohesion, which encourages an agent to remain close to the group. Despite the simplicity of these rules, they are able to reproduce realistic group motion similar to bird flocks and fish schools. This work provided an important foundation for later studies and inspired many computational models of collective behavior \citep{Hartman2006,Vicsek2012}.

Alongside algorithmic approaches, another major direction consists of describing collective behaviors through dynamical models. While algorithms prescribe rules that individual agents must follow, dynamical modeling focuses on the time evolution of the system and on the mathematical laws governing the interactions between agents \citep{Carrillo2010}. This approach is useful for studying the emergence of ordered motion, the transition from disorder to coordination, and the conditions under which a group of agents can organize itself.

One of the major contribution in this direction was the model proposed by Vicsek and collaborators for self-propelled particles \citep{Vicsek1995}. In the Vicsek model, particles move in a continuous space and update their direction according to the average direction of nearby particles. This model revealed a transition from a disordered state, where the particles move in different directions, to an ordered state, where they move coherently in a common direction. Such directional ordering is closely related to the notion of \emph{flocking}, which refers to the synchronization of the directions of motion of agents in a group.

The idea of synchronization is much broader than flocking and appears in many natural and artificial systems. Early works by Winfree on biological oscillators and by Kuramoto on populations of coupled nonlinear oscillators laid the foundations of modern synchronization theory \citep{Winfree1967,Kuramoto1975}. Since then, synchronization has been used to explain a wide range of phenomena, including biological rhythms, flashing of fireflies, collective clapping, neuronal activity, and power-grid dynamics \citep{Pikovsky2001,Buck1988,Neda2000,Womelsdorf2007,Rohden2012}. However, in many classical studies of synchronization, the spatial positions of the oscillators are not explicitly considered.

In parallel with synchronization theory, several models have focused mainly on the spatial organization of agents. These studies investigate aggregation, pattern formation, and swarming dynamics \citep{Bernoff2013,Topaz2006,Topaz2004}. In many swarming models, agents interact through attraction and repulsion forces. Attraction helps maintain the cohesion of the group, whereas repulsion prevents collisions and avoids overcrowding. The balance between these two effects can generate stable spatial structures and collective aggregation patterns. This spatial organization is a key feature of \emph{swarming}.

More recently, researchers have started to investigate systems in which spatial motion and synchronization of internal variable are coupled \cite{sar2026interplay}. In such systems, the movement of agents can influence their synchronization, while their internal states can also affect their spatial organization. Studies on mobile agents have shown that mobility can significantly modify the synchronization properties of coupled dynamical systems \citep{Frasca2008,Chowdhury2019,Majhi2019}. This has led to the development of models that combine both spatial degrees of freedom and internal phase dynamics. Agents that simultaneously swarm in space and synchronize their internal phases are now commonly referred to as \emph{swarmalators} \cite{o2017oscillators,sar2022dynamics,Sar2023Pinning,Sar2022Competitive,o2022collective,yoon2022sync,o2022swarmalators,hao2023attractive,sar2023swarmalators,anwar2024forced,anwar2025forced,lizarraga2023synchronization,o2026time,blum2024swarmalators,lambu2026delay,sar2025effects,hong2023swarmalators,lr2r-ynzs,anwar2024collective,senthamizhan2026swarmalators,sar2025strategy,o2025global,o2025stability,o2026unsteady,ghosh2026emergent,kongni2023phase,kongni2024expected,smith2024swarmalators,sar2026interplay,djeudjo2026role}

Despite these developments, most swarmalator models still describe the internal variable as a generic oscillator, most often of Kuramoto type. A notable example is the work in~\cite{GPKG2024}, where the internal dynamics was modeled using a R\"ossler system. Those assumptions may be less appropriate when the internal variable represents a  quantity associate to a social value, such as an opinion. Here, we thus decided to follow a different route, the internal variable is defined to be an opinion whose dynamics is governed by a continuous-time analogue of the Deffuant model~\cite{deffuant2000mixing}.

In the original Deffuant model, opinions are real numbers in the interval $[0,1]$. At each discrete time step, two agents are randomly selected from the population and compare their opinions. If the opinion difference is smaller than a confidence threshold $d\in(0,\frac{1}{2})$, i.e., $|O_i(t)-O_j(t)|<d$, the two agents move toward a local compromise:
\begin{equation*}
\begin{aligned}
O_i(t+1)
&= O_i(t)+\mu\bigl(O_j(t)-O_i(t)\bigr),\\
O_j(t+1)
&= O_j(t)+\mu\bigl(O_i(t)-O_j(t)\bigr)\,.
\end{aligned}
\end{equation*}
The parameter $\mu\in(0,\frac{1}{2})$ controls the convergence rate toward compromise. If $\mu=1/2$, the two agents immediately reach their average opinion. If $|O_i(t)-O_j(t)|>d$, the agents do not influence each other, hence $O_i(t+1)
= O_i(t)$ and $O_j(t+1)
= O_j(t)$. An analogous version of the above model, can be defined in the case of continuous time
\begin{equation}
    \label{eq:conttD}
    \frac{dO_i(t)}{dt}=\frac{\mu}{N}\sum_j\Theta\left(d-|O_j(t)-O_i(t)|\right) \left(O_j(t)-O_i(t)\right)\, ,
\end{equation}
where $\Theta(s)$ is the Heaviside function, i.e., $\Theta(s)=0$ if $s<0$ and $\Theta(s)=1$ if $s\geq 0$, and $N$ is the population size.

The goal of the present work is to combine swarming dynamics with opinion dynamics, described by the Deffuant model, in such a way that the two dynamics influence each other, as in the swarmalator framework. As will be shown in the following, our results indicate that the confidence threshold controls the number of opinion groups, while the strength of the opinion-dependent spatial attraction determines how these groups are organized in space. More precisely, this parameter allows us to control whether opinion groups spatially merge or remain separated. In addition, in the full-consensus state, where all agents share the same opinion and organize themselves in a disk, we use a semi-analytical approach to derive the radius of the stationary swarm when the nonlinear attraction kernel is used. We also explore the dynamics in the case of the linear kernel (see Appendix~\ref{app:linear_kernel}), and observe that the conclusions are similar to those obtained for the nonlinear kernel. 

The rest of this work is organized as follows: in Sec.~\ref{sec:socswarm}, we present the model on which this study is based, as well as the characterization parameters used throughout the work; in Sec.~\ref{sec:numres}, we present the numerical results obtained; in Sec.~\ref{sec:consstate}, we explain how the radius of the consensus swarm is computed and show the agreement between our analytical results and numerical simulations; finally, in Sec.~\ref{sec:conclusion}, we  conclude.

\section{The social swarm model}
\label{sec:socswarm}

The aim of this section is to introduce the {\em social swarm model} , i.e., where agents possess an internal state represented by a variable encoding for an opinion, and they move in a two-dimension plane; moreover the two dynamics are coupled and thus influence each other. The model is inspired by the swarmalator framework, in which the spatial motion of agents is coupled with the evolution of an internal state. However, in contrast with classical swarmalator models \cite{sar2026interplay}, where the internal variable is usually an oscillatory phase, the internal state considered here represents a social opinion governed by a Deffuant-type bounded-confidence dynamics. Let us observe that other models could be used to describe the evolution of the opinion state. The main idea upon which the model is rooted, is that spatial and social interactions mutually influence each other: agents with similar opinions experience stronger spatial attraction, whereas agents with very different opinions interact more weakly in physical space. Conversely, opinion dynamics is affected by spatial proximity: nearby agents interact more strongly in opinion space than distant agents, promoting faster opinion convergence among spatially close individuals. This coupling between collective motion and opinion dynamics leads to the following equations for a system composed by $N$ agents:
\begin{widetext}
\begin{eqnarray}
\dfrac{d x_i}{dt} &=& \frac{1}{N} \sum_{j \neq i} \left[ \frac{x_j - x_i}{\sqrt{(x_j - x_i)^2 + (y_j - y_i)^2}} (A + J e^{-\lambda |O_j - O_i|}) - B \frac{x_j - x_i}{(x_j - x_i)^2 + (y_j - y_i)^2}  \right],\label{eq:sociswarmx}\\
\dfrac{d y_i}{dt} &=& \frac{1}{N} \sum_{j \neq i} \left[ \frac{y_j - y_i}{\sqrt{(x_j - x_i)^2 + (y_j - y_i)^2}} (A + J e^{-\lambda |O_j - O_i|}) - B \frac{y_j - y_i}{(x_j - x_i)^2 + (y_j - y_i)^2}   \right],\label{eq:sociswarmy}\\
\dfrac{d O_i}{dt} &=& \frac{\mu}{N} \sum_{j \neq i} \frac{ \Theta(d - | O_j - O_i|) (O_j - O_i)}{\sqrt{(x_j - x_i)^2 + (y_j - y_i)^2}}\label{eq:sociswarmo}\, .
\end{eqnarray}
\end{widetext}
where $(x_i,y_i)\in\mathbb{R}^2$ describe the spatial position of the $i$-th agent, whereas $O_i\in [0,1]$ the opinion of the same agent. Equations ~\eqref{eq:sociswarmx} and~\eqref{eq:sociswarmy} recall the classical swarmalator model~\cite{o2017oscillators} where the cosine term, depending on the internal phase, is now replaced by a function of the difference of opinion variables. The last Eq.~\eqref{eq:sociswarmo} has been directly obtained form the continuous time version of the Deffuant model dynamics~\eqref{eq:conttD}, with the inclusion of a feedback term depending on the distance among agents: the farther the agents, the smaller the (possible) change in their opinions.

The parameter $d\in (0,1)$ represents the bounded-confidence threshold, and $\mu$ sets the time scale of the opinion dynamics. The parameter $\lambda\geq0$ controls the feedback of opinions on spatial attraction. The model contains thus six parameters: $A$, $B$, $J$, $\lambda$, $d$, and $\mu$. In the numerical results below, unless otherwise stated, we set $A=1$ and $B=1$ and focus on the parameters most directly related to opinion dynamics and opinion-space feedback.

Before to proceed with the analysis of the model, let us define some relevant metrics that will be used for the study. To distinguish between stationary and active states, we introduce the {\em average swarm velocity} at time $t$, $V(t)$, defined as
\begin{equation}
\label{eq:avevel}
V(t) = \frac{1}{N}\sum_{i=1}^{N} \|v_i(t)\|\, , \quad
\text{where}\quad
\|v_i(t)\|
= \sqrt{\dot{x}_i^{\,2}(t)+\dot{y}_i^{\,2}(t)}\, .
\end{equation}
 When $V(t)$ converges to a value close to zero after a transient, the swarm is essentially stationary. In contrast, a persistent positive value indicates that the agents keep moving.

\begin{figure*}[ht!]
        \includegraphics[width=0.6\textwidth]{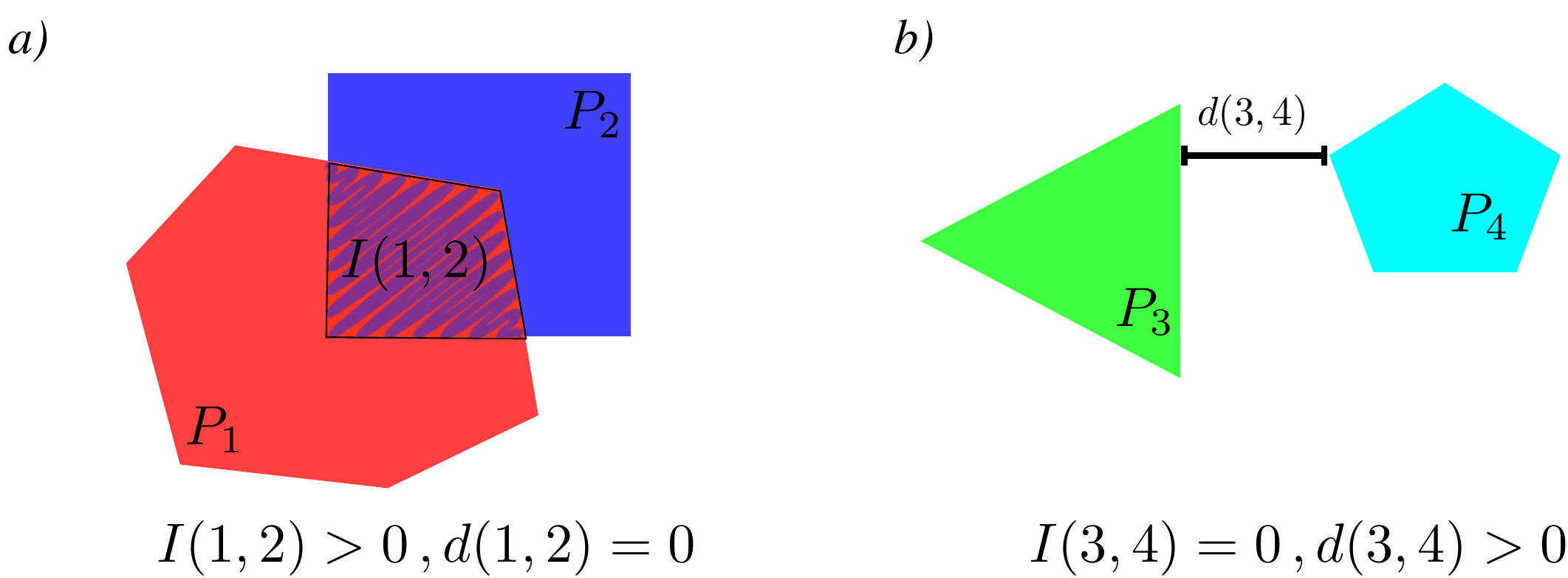}
        \caption{{\bf Overlap and distance metrics}. Panel (a) shows two generic polygons, $P_1$ (red) and $P_2$ (blue), with a nonempty intersection (dashed region), so that $0<I(1,2)<1$ and the distance between them is zero. Panel (b) shows two disjoint polygons, $P_3$ (green) and $P_4$ (cyan), for which $I(3,4)=0$ and $d(3,4)>0$.}
    \label{fig:figIntdist}
\end{figure*}

The number of opinion groups will be denoted by $N_g$, and are defined by the fact that agents in the same group share the same asymptotic opinion. To determine the spatial arrangement of those groups, we define two metrics, the {\em overlap} and the {\em distance}. More precisely, let us suppose that, after a sufficiently long transient time, the population is divided into $N_g$ opinion groups, for each group $\ell=1,\dots, N_g$, we collect the spatial positions of its agents and construct the corresponding convex hull, denoted by $P_\ell\subset\mathbb{R}^2$. The overlap between two spatial groups is measured by
\begin{equation}
\label{eq:Imat}
I(i,j) = \frac{\text{Area}(P_i \cap P_j)}{\min\left(\text{Area}(P_i), \text{Area}(P_j)\right)} \quad \forall i,j\in\{1,\dots,N_g\}\, ,
\end{equation}
where $\text{Area}(P)$ denotes the area of the polygon $P$. If one polygon is fully contained in the other, then $I(i,j)=1$. If the two polygons are disjoint, then $I(i,j)=0$ (\textcolor{blue}{see figure \ref{fig:figIntdist}} (b)). Therefore $I(i,j)\in[0,1]$. If a group contains a single agent, the denominator is set to one in order to avoid division by zero.
To quantify the separation between two opinion groups, we define the distance
\begin{equation}
\label{eq:distmat}
d(i,j) = \{\text{Euclidean distance between $P_i$ and $P_j$}\}\, .
\end{equation}
If the two polygons overlap, or if one polygon is contained in the other, we set $d(i,j)=0$ (\textcolor{blue}{see figure \ref{fig:figIntdist}} (a)).

\section{Numerical results}
\label{sec:numres}

\begin{figure*}[ht!]
    \centering
    \setlength{\tabcolsep}{0pt}
    \begin{tabular}{ccccc}
        \includegraphics[width=0.2\textwidth]{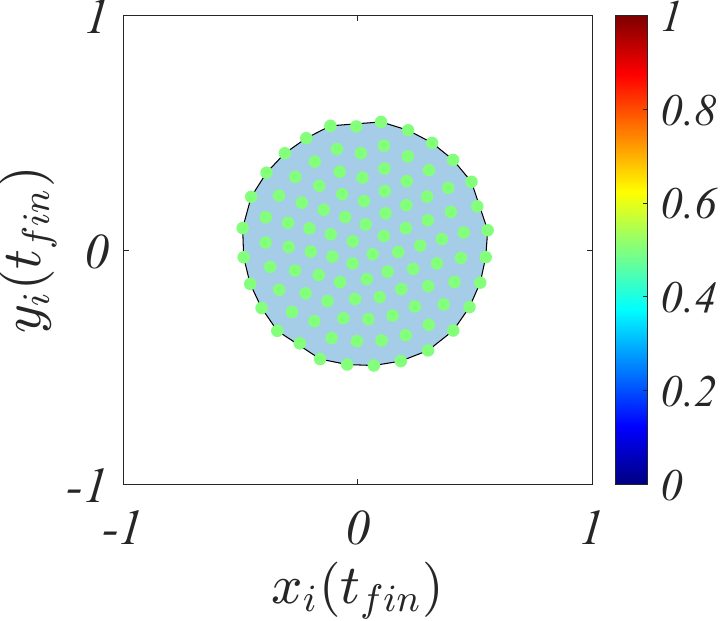}  &
        \includegraphics[width=0.2\textwidth]{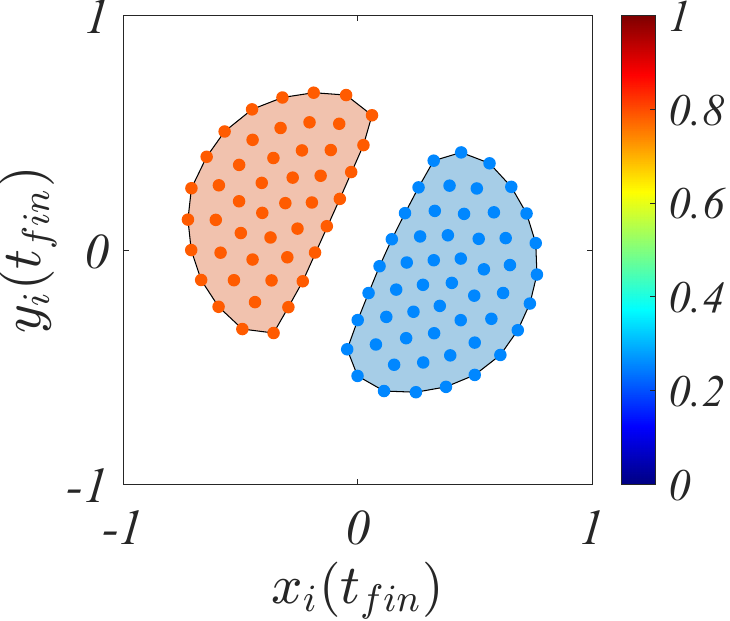} &
        \includegraphics[width=0.2\textwidth]{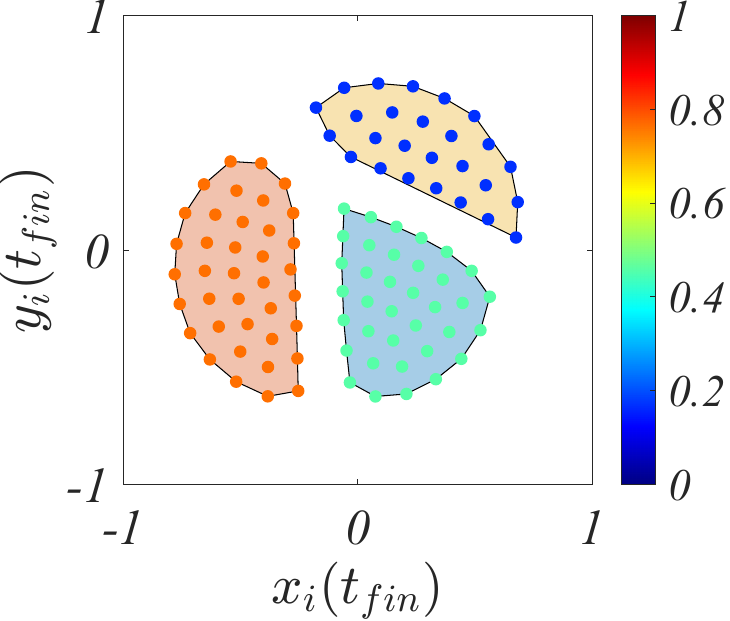}  &
        \includegraphics[width=0.2\textwidth]{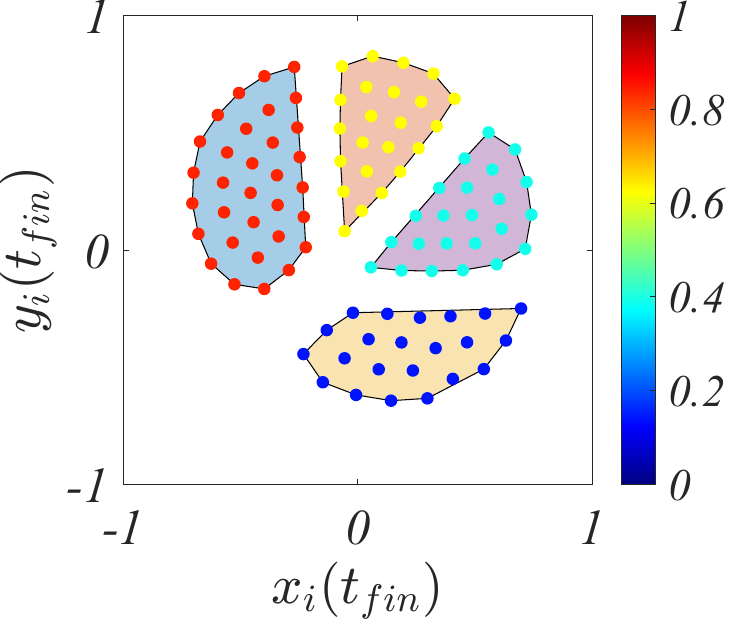} &
        \includegraphics[width=0.2\textwidth]{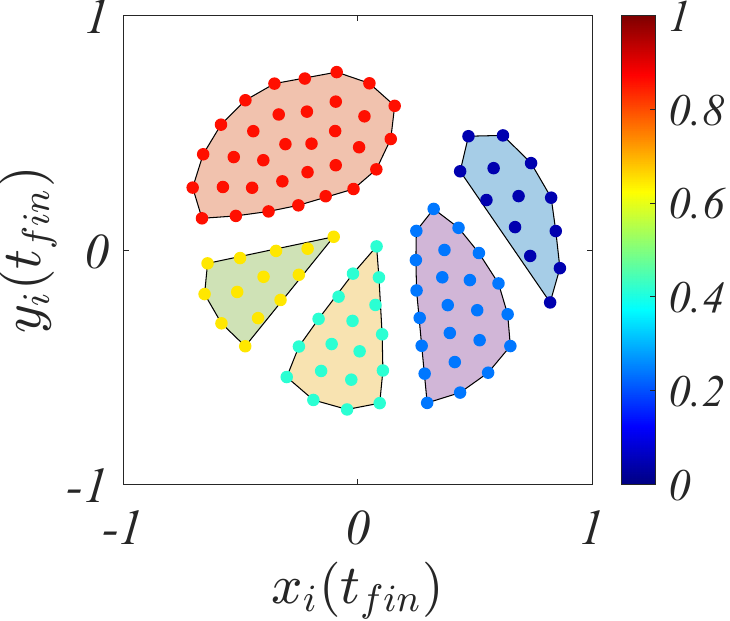}\\ 
        \textbf{(a1)} & \textbf{(b1)} & \textbf{(c1)} & \textbf{(d1)} & \textbf{(e1)} \\

        \includegraphics[width=0.2\textwidth]{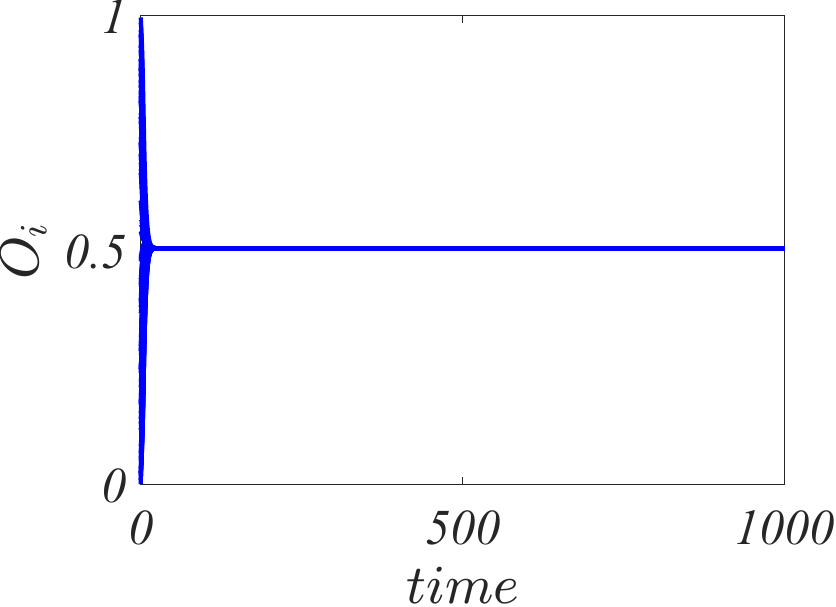}&
        \includegraphics[width=0.2\textwidth]{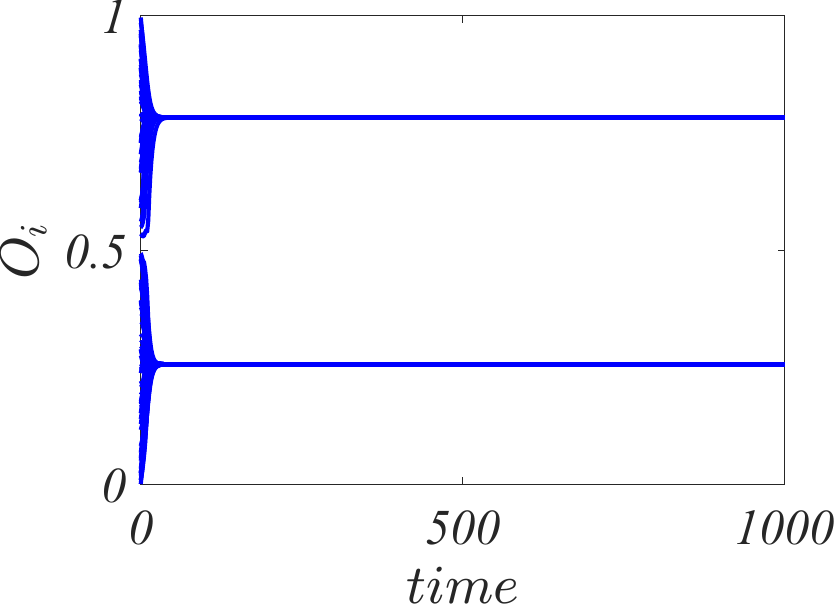} &
        \includegraphics[width=0.2\textwidth]{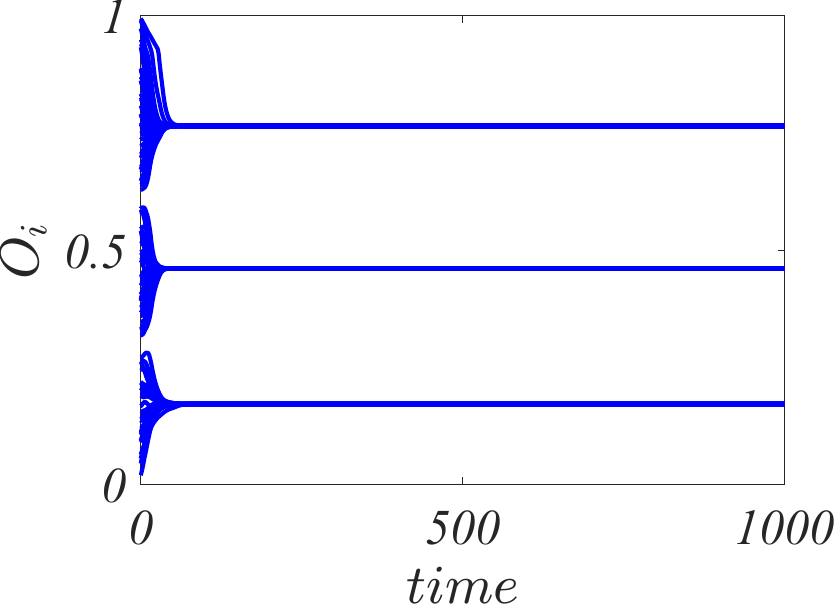}  &
        \includegraphics[width=0.2\textwidth]{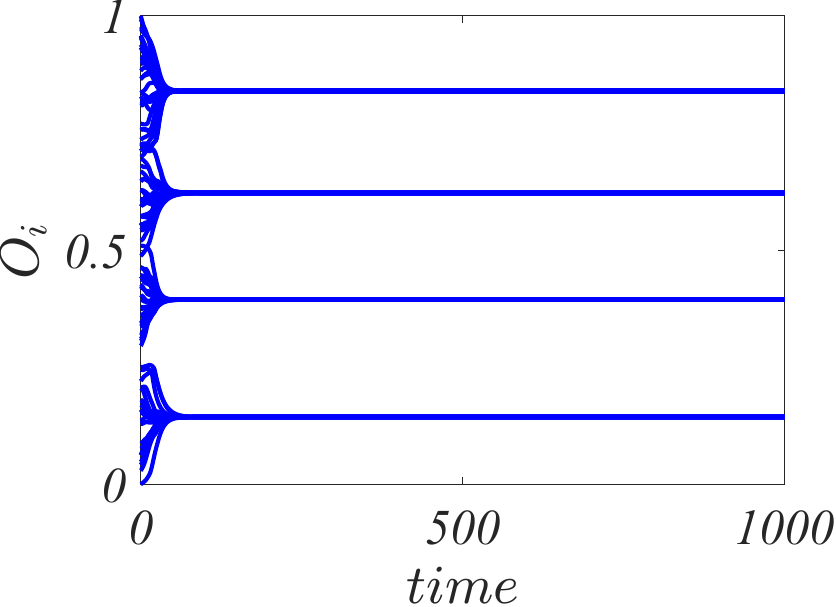} &
        \includegraphics[width=0.2\textwidth]{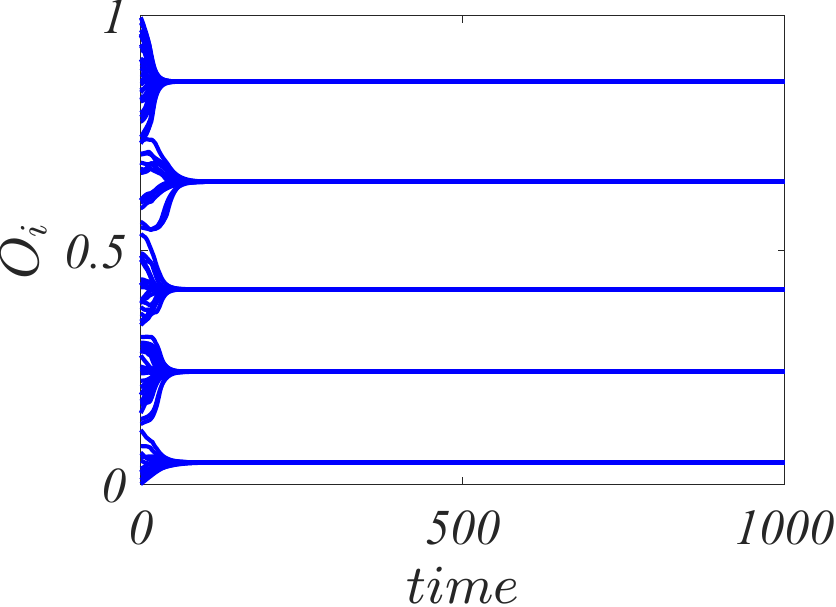}\\ 
        \textbf{(a2)} & \textbf{(b2)} & \textbf{(c2)} & \textbf{(d2)} & \textbf{(e2)} \\

        \includegraphics[width=0.2\textwidth]{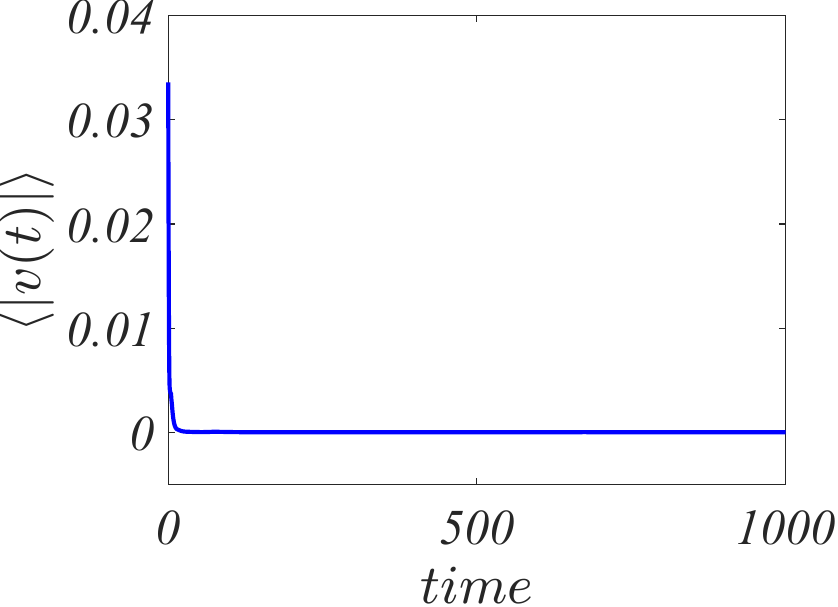}&
        \includegraphics[width=0.2\textwidth]{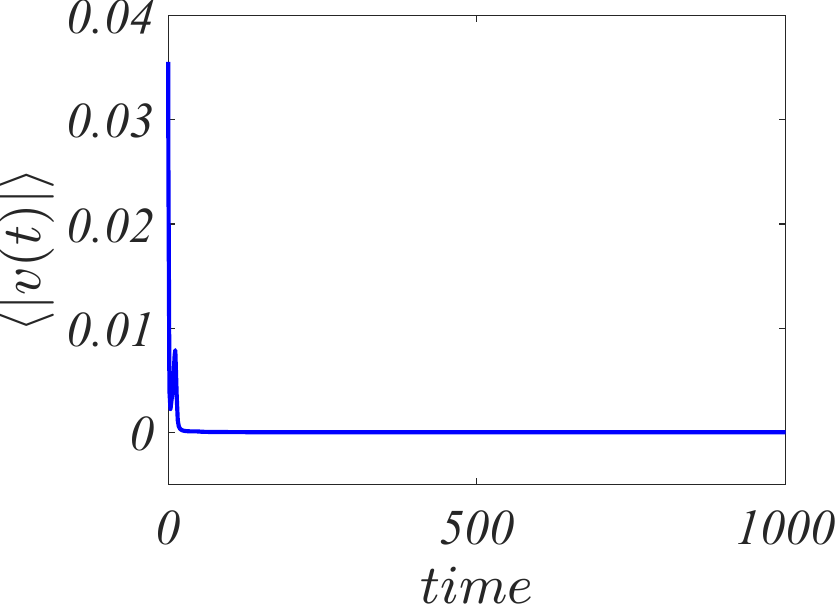} &
        \includegraphics[width=0.2\textwidth]{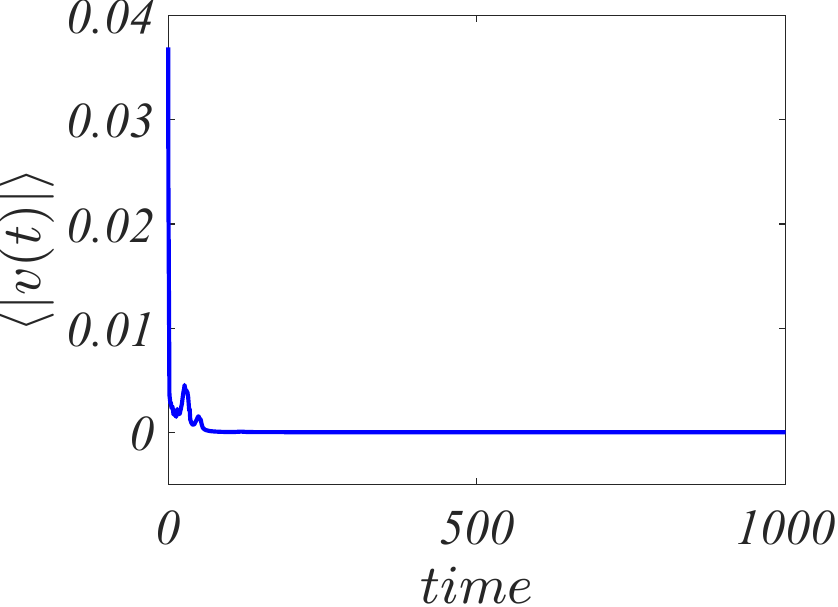}  &
        \includegraphics[width=0.2\textwidth]{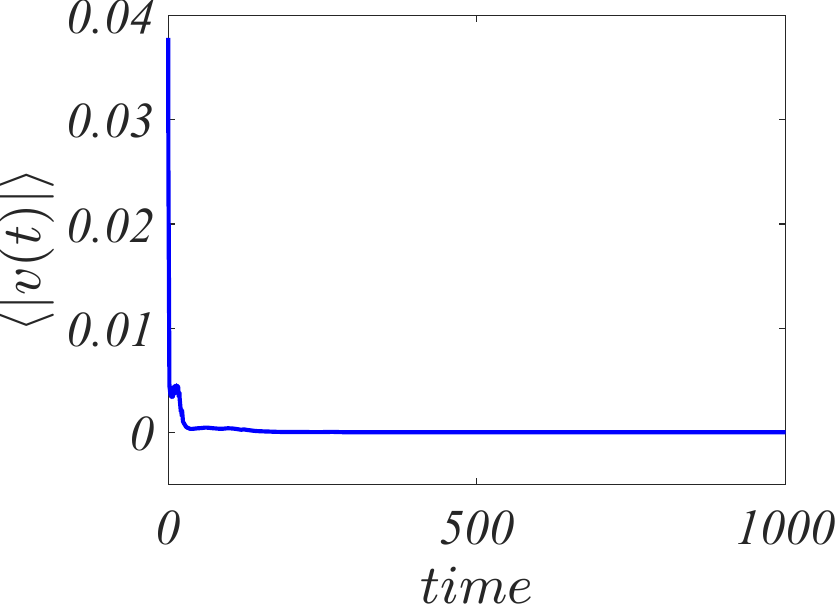} &
        \includegraphics[width=0.2\textwidth]{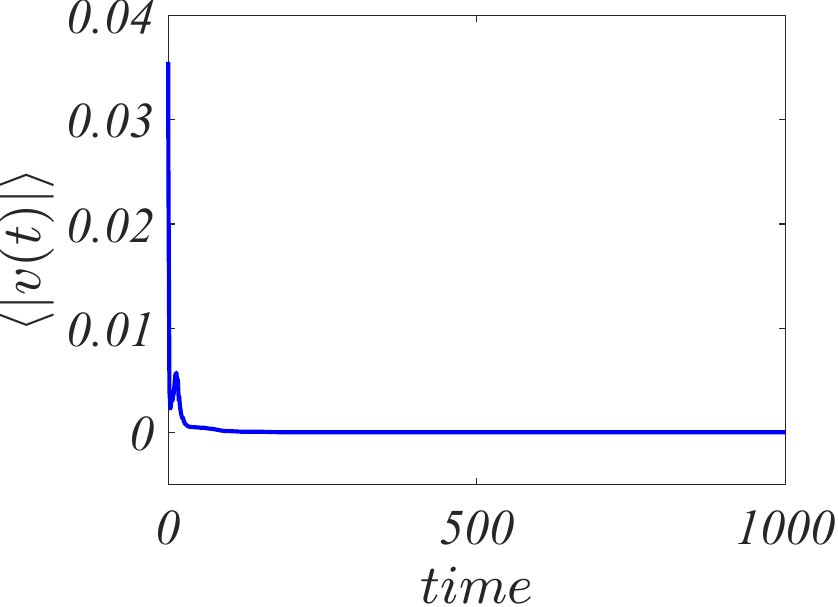}\\ 
        \textbf{(a3)} & \textbf{(b3)} & \textbf{(c3)} & \textbf{(d3)} & \textbf{(e3)} \\
    \end{tabular}
    \caption{{\bf Effect of bounded-confidence threshold, $d$, on the social swarmalator dynamics}. (a) $d=\frac{1}{2}$, (b) $d=\frac{1}{4}$, (c) $d=\frac{1}{6}$, (d) $d=\frac{1}{8}$, and (e) $d=\frac{1}{10}$. The top row (a1--e1) shows the final configuration in the $x-y$ plane; the middle row (a2--e2) shows the opinion dynamics; and the bottom row (a3--e3) shows the average velocity. The remaining parameters are $J=1.0$, $N=100$, $\mu=0.2$, and $\lambda=3$.}
    \label{fig:fig1}
\end{figure*}

To unravel possible new features of the social swarm model due to the opinion dynamics, we start the numerical analysis by first studying the role of the bounded-confidence threshold $d$. We thus fix $J=1$, $\mu=0.2$, and $\lambda=3$, and we compare the model outcomes for several values of $d$. The results are shown in Fig.~\ref{fig:fig1}, from left to right, the columns correspond to decreasing values of $d$, i.e.,  $d=\frac{1}{2}$, $d=\frac{1}{4}$, $d=\frac{1}{6}$, $d=\frac{1}{8}$, and $d=\frac{1}{10}$; top row displays the spatial configuration, with agents colored according to their final opinion and convex hulls indicating the spatial extent of each opinion group. The middle row shows the time evolution of the opinions, and the bottom row shows the average swarm velocity.

The numerical results clearly show that $d$ controls the number of opinion clusters. In agreement with the bounded-confidence mechanism, the number of groups satisfies approximately $N_g\leq \frac{1}{2d}$ (see middle row in Fig.~\ref{fig:fig1}). For the parameters used to obtain the results shown in Fig.~\ref{fig:fig1}, different opinion groups are also spatially separated (see top row). Hence, in this regime, changing $d$ controls both the number of opinion clusters and the number of spatial clusters. The average velocity decays to zero in all cases, indicating that the final configurations are stationary (see bottom row).

\begin{figure*}[ht!]
    \centering
    \setlength{\tabcolsep}{0pt}
    \begin{tabular}{cccc}
    \includegraphics[width=0.22\textwidth]{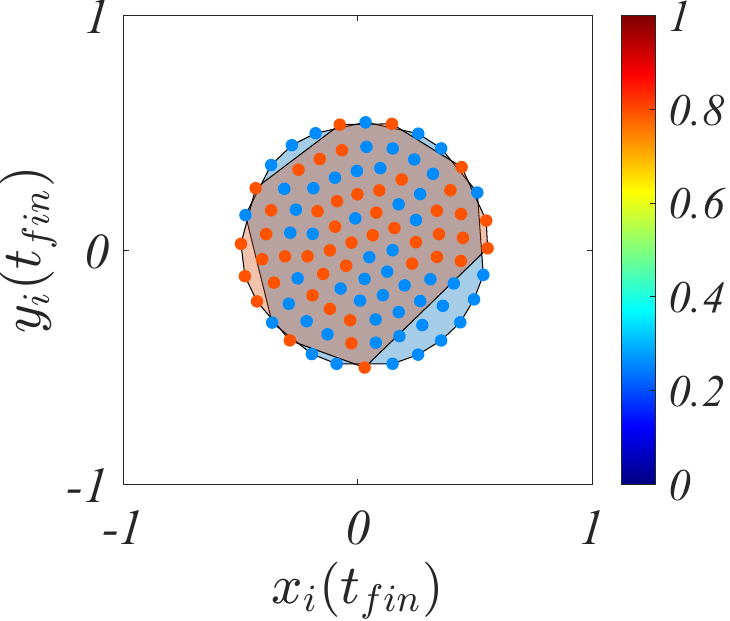} &
        \includegraphics[width=0.22\textwidth]{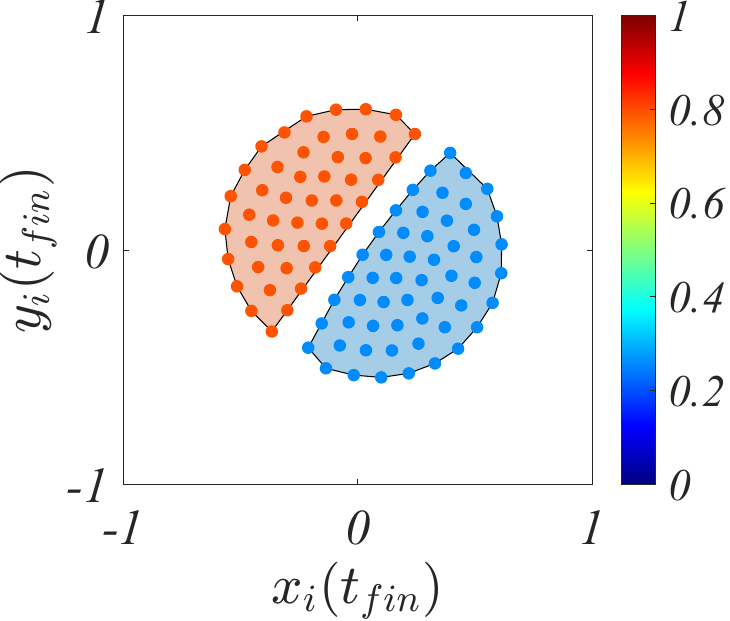} &
        \includegraphics[width=0.22\textwidth]{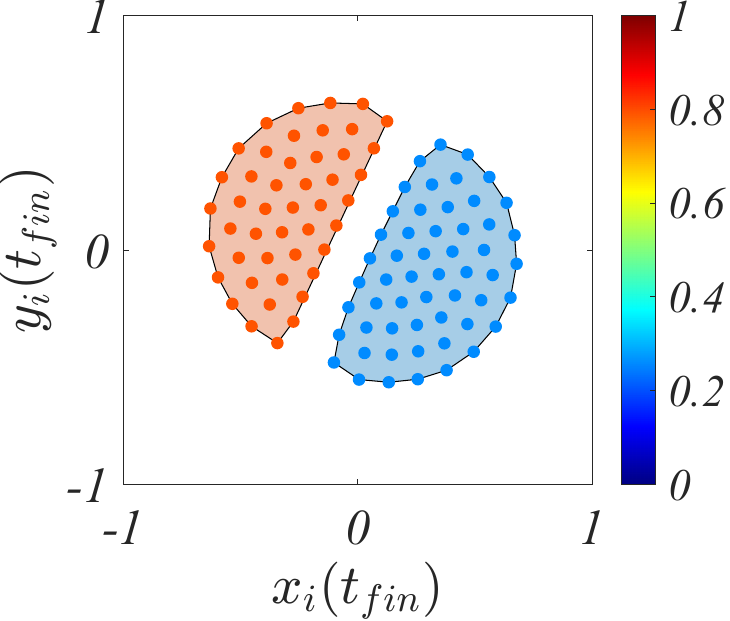}  &
        \includegraphics[width=0.22\textwidth]{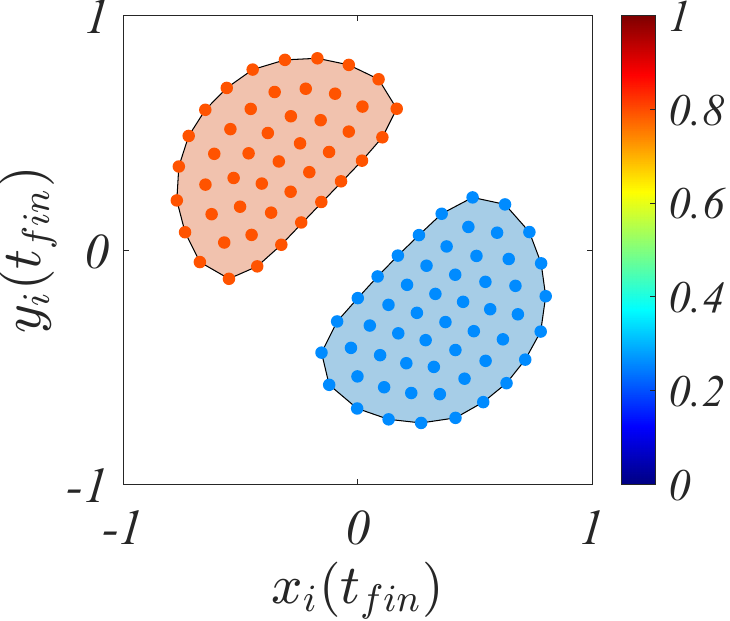}\\ 
        \textbf{(a1)} & \textbf{(b1)} & \textbf{(c1)} & \textbf{(d1)}\\
        \includegraphics[width=0.22\textwidth]{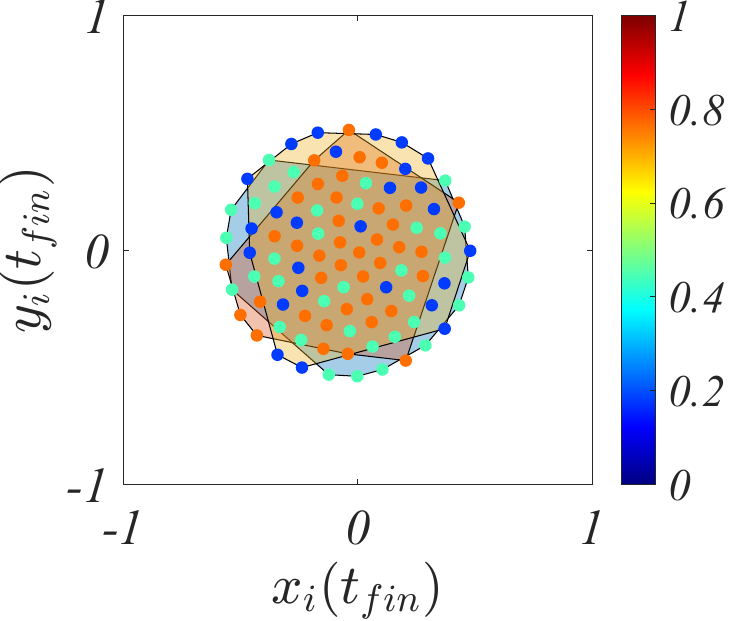} &
        \includegraphics[width=0.22\textwidth]{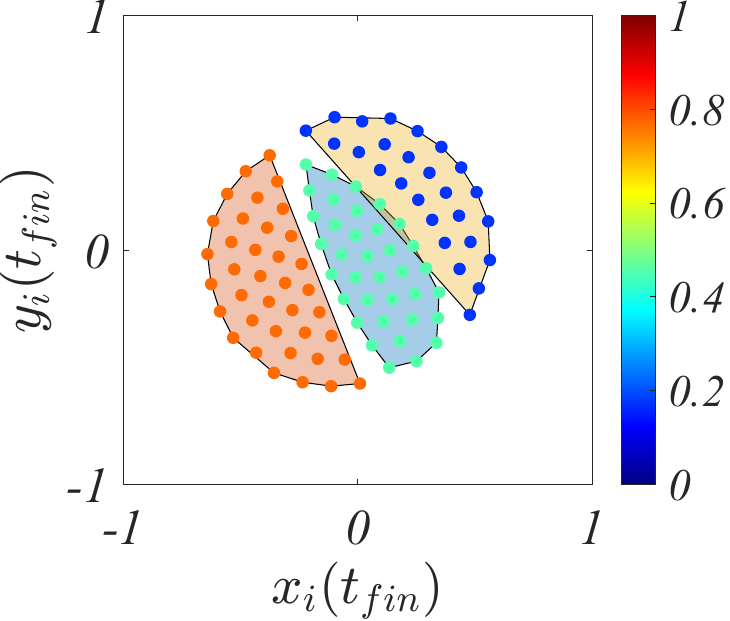} &
        \includegraphics[width=0.22\textwidth]{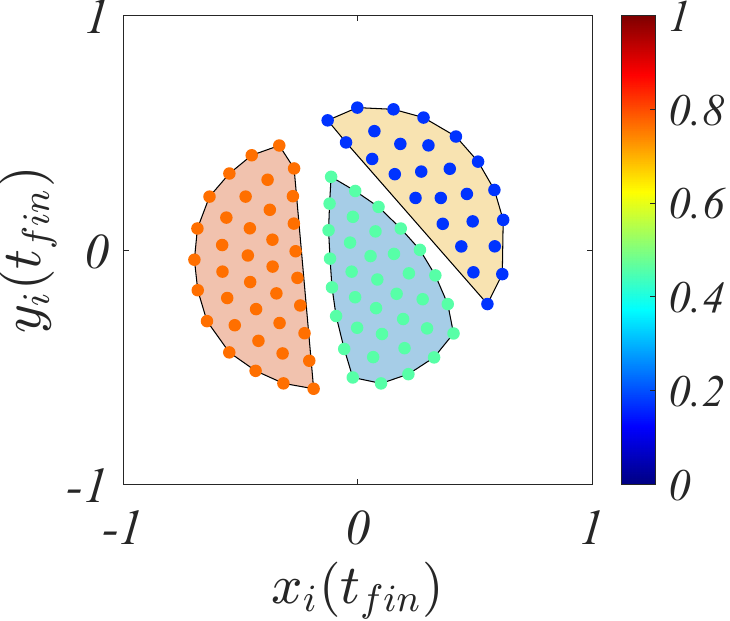}  &
        \includegraphics[width=0.22\textwidth]{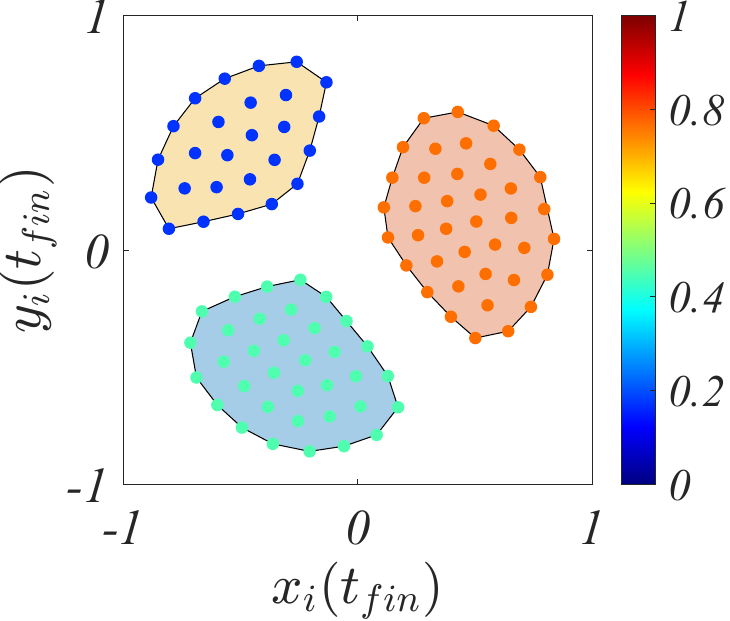}\\ 
        \textbf{(a2)} & \textbf{(b2)} & \textbf{(c2)} & \textbf{(d2)}\\
        \includegraphics[width=0.22\textwidth]{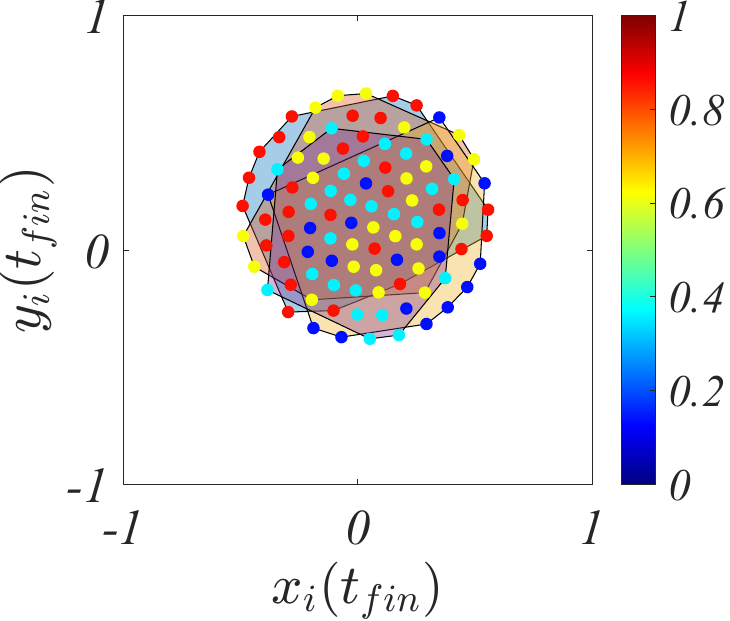} &
        \includegraphics[width=0.22\textwidth]{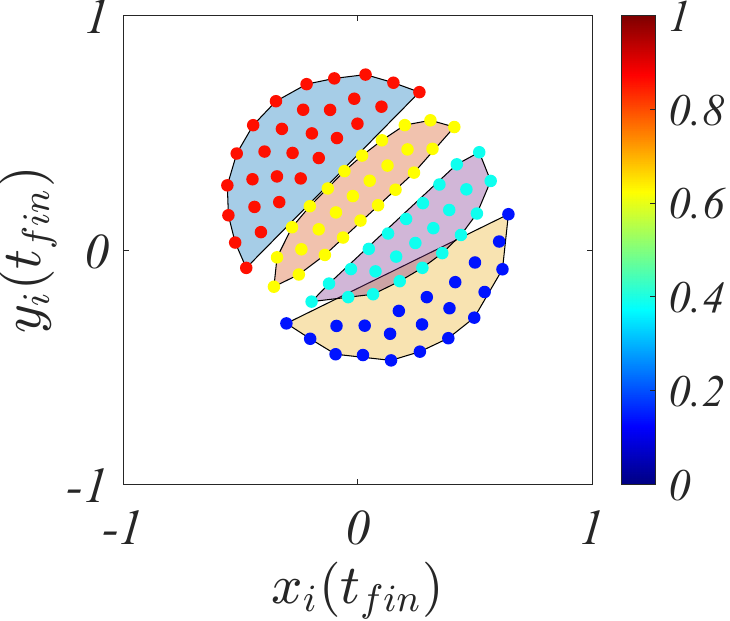} &
        \includegraphics[width=0.22\textwidth]{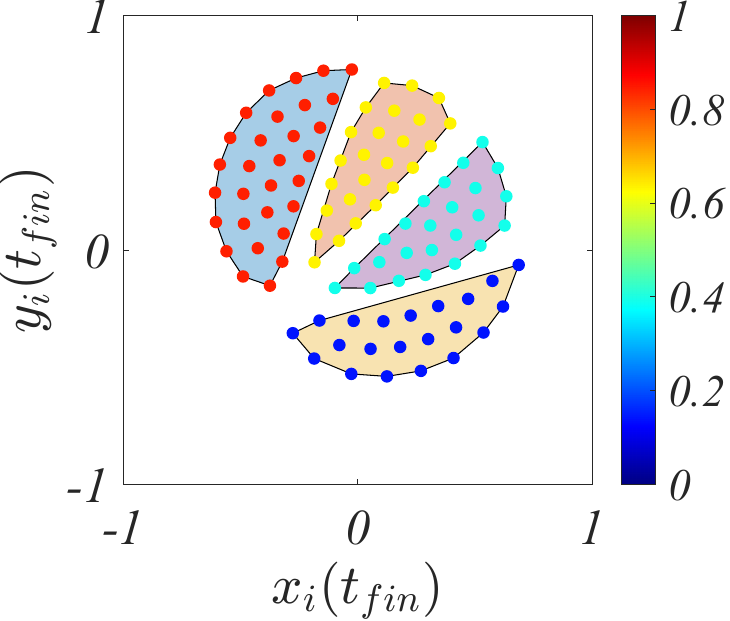}  &
        \includegraphics[width=0.22\textwidth]{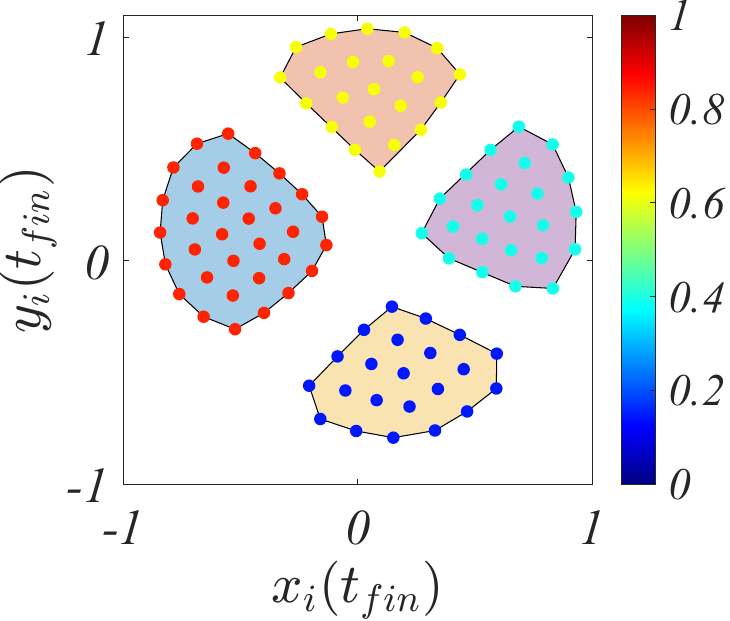}\\ 
        \textbf{(a3)} & \textbf{(b3)} & \textbf{(c3)} & \textbf{(d3)}\\
        \includegraphics[width=0.22\textwidth]{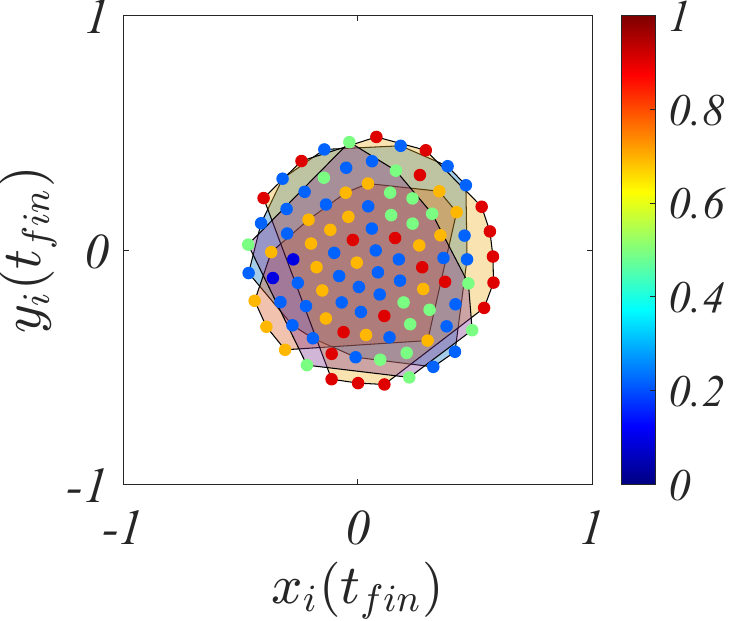} &
        \includegraphics[width=0.22\textwidth]{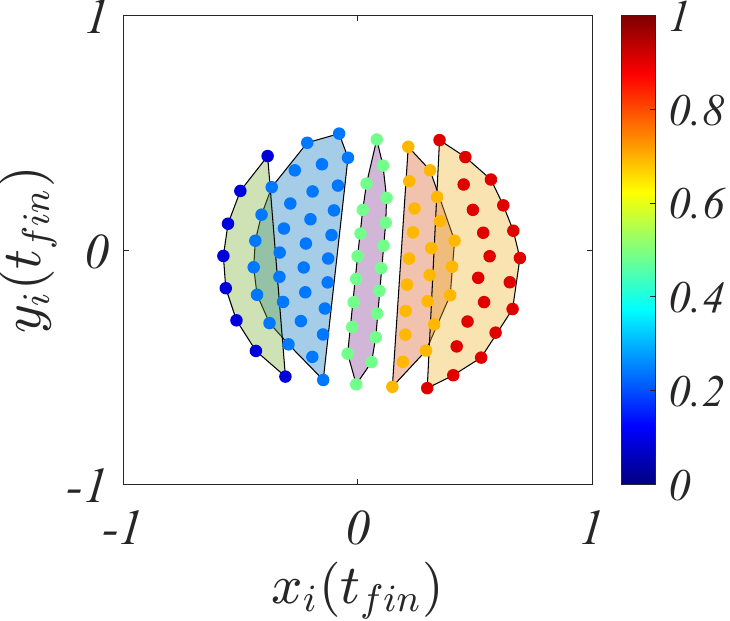} &
        \includegraphics[width=0.22\textwidth]{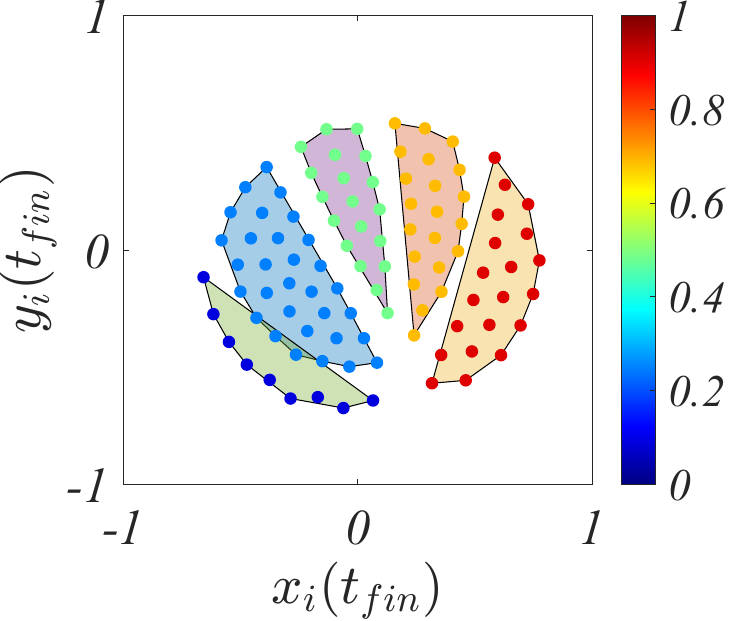}  &
        \includegraphics[width=0.22\textwidth]{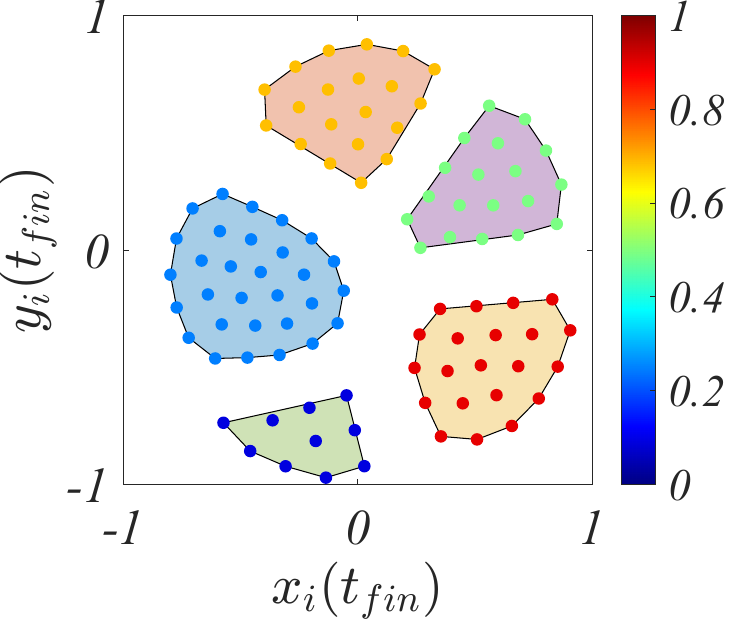} \\ 
        \textbf{(a4)} & \textbf{(b4)}  & \textbf{(c4)} & \textbf{(d4)}\\
    \end{tabular}
    \caption{{\bf Effect of $d$ and $\lambda$}. The rows correspond, from top to bottom, to $d=\frac{1}{4}$, $d=\frac{1}{6}$, $d=\frac{1}{8}$, and $d=\frac{1}{10}$. The columns correspond, from left to right, to $\lambda=0.05$, $\lambda=0.8$, $\lambda=1.5$, and $\lambda=10$. The remaining parameters are $J=1.0$, $N=100$, and $\mu=0.2$.}
    \label{fig:fig2}
\end{figure*}

The next step consists in studying the effect of $\lambda$, which controls how strongly opinion differences affect spatial attraction. We thus fix $J=1$ and $\mu=0.2$ and vary both $\lambda$ and $d$. The results are reported in Fig.~\ref{fig:fig2}. The columns correspond (from left to right) to $\lambda=0.05$, $\lambda=0.8$, $\lambda=1.5$, and $\lambda=10$, whereas the rows correspond to (from top to bottom) $d=\frac{1}{4}$, $d=\frac{1}{6}$, $d=\frac{1}{8}$, and $d=\frac{1}{10}$. One can observe that as before, the value of $d$ determines the number of opinion groups whereas the parameter $\lambda$ mainly controls their spatial arrangement. For small $\lambda$, the exponential factor $e^{-\lambda |O_j-O_i|}$ remains close to one even for agents with different opinions, and distinct opinion groups can remain spatially close (see left and second to left columns). As $\lambda$ increases, agents with different opinions attract each other less strongly, and the corresponding spatial groups become more clearly separated (see right and second to right columns). In all cases, the groups are static, indeed the average group velocity is zero (data not shown).

\begin{figure*}[ht!]
    \centering
    \setlength{\tabcolsep}{0pt}
    \begin{tabular}{ccccc}
        \includegraphics[width=0.32\textwidth]{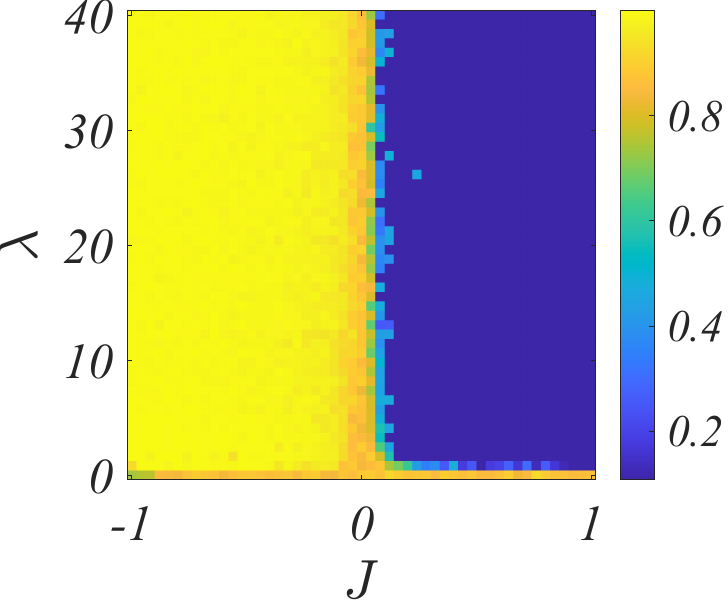}  &
        \includegraphics[width=0.32\textwidth]{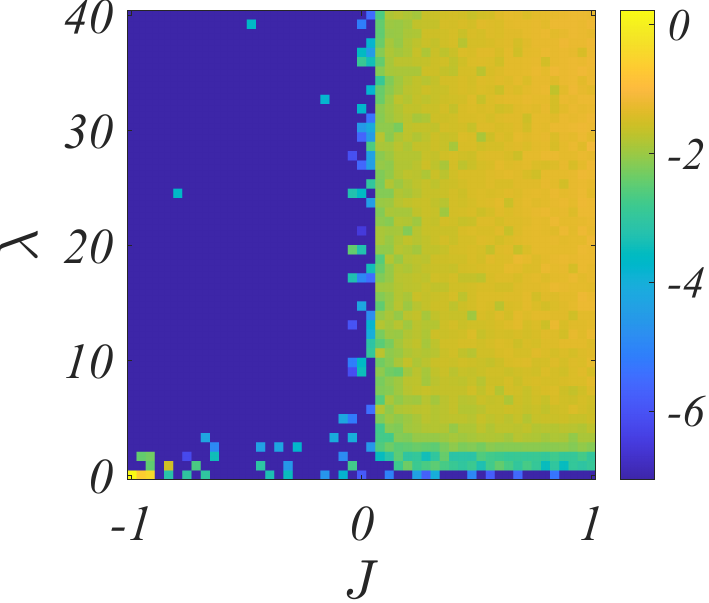} &
        \includegraphics[width=0.33\textwidth]{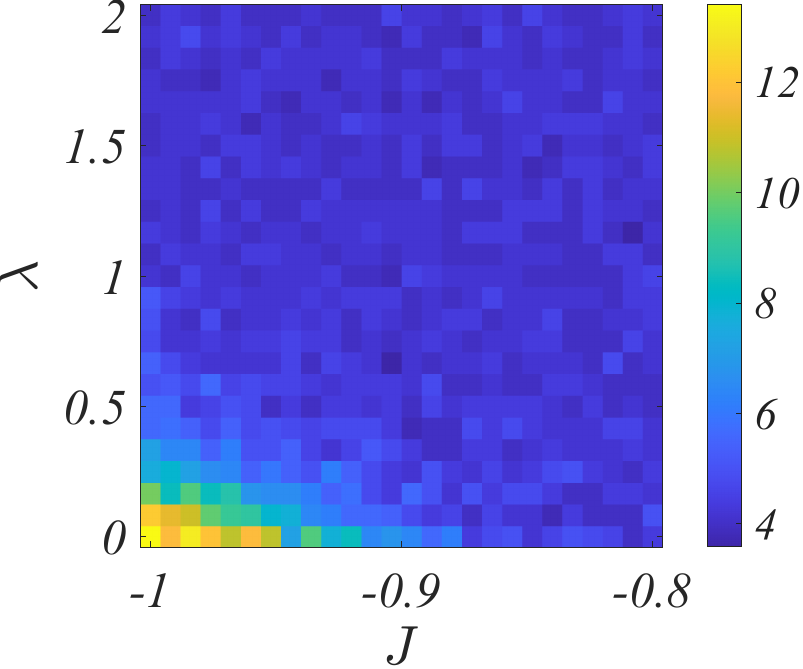}\\
       \textbf{(a)} & \textbf{(b)} & \textbf{(c)} 
    \end{tabular}
    \caption{{\bf Effect of $J$ and $\lambda$.} The left panel (a) shows the average group overlap using a color scale: yellow corresponds to large values, close to one, indicating almost completely overlapping groups, whereas blue corresponds to well-separated groups. The middle panel (b) shows the logarithm of the average distance between groups. In the region where groups completely overlap, this logarithm is not defined. For positive $J$, groups separate from one another, and their distance increases when both $J$ and $\lambda$ are large. The right panel (c) shows the average number of opinion groups at equilibrium. We restrict the parameter ranges to $0 \leq \lambda \leq 2$ and $-1 \leq J \leq -0.8$ in order to highlight the behavior near $\lambda \sim 0$ and $J \sim -1$; moreover the number of groups remains almost constant for larger values of $J$ and $\lambda$, with a value close to $5$ (data not shown). The remaining parameters are set to $d=1/10$, $N=100$, and $\mu=0.2$.}
    \label{fig:fig3}
\end{figure*}

We then consider the combined effect of $J$ and $\lambda$. We fix $\mu=0.2$, $d=\frac{1}{10}$, and $N=100$, and simulate the model for $\lambda\in[0,40]$ and $J\in[-1,1]$. The value of the bounded confidence threshold has been set small enough for several opinion clusters to emerge. For each parameter pair, we compute the average group overlap, $\sum_{ij} I(i,j)/N^2$, the average group distance, $\sum_{ij} d(i,j)/N^2$, and the average number of opinion groups. The results are shown in Fig.~\ref{fig:fig3}. For positive $J$, opinion groups tend to form distinct spatial clusters, especially when $\lambda$ is sufficiently large; this claim can be appreciated by observing that groups do not overlap (see blue region in the left panel) and thus their relative distance is positive (see olive-green and orange regions in the middle panel). In contrast, for negative $J$, the spatial organization is less stable and the overlap between opinion groups is typically larger (see yellow region in the left panel). This difference is caused by the attraction term $A+J e^{-\lambda |O_j-O_i|}$, which can become very small when $J<0$.

Since $d=\frac{1}{10}$, the bounded-confidence rule suggests that the typical number of opinion groups should be close to $\frac{1}{2d}=5$. This is indeed observed in a large portion of the parameter space. However, for negative $J$ and small values of $\lambda$, the system can produce many more opinion groups. In that regime, the spatial motion can push agents far apart, which weakens the opinion interaction in Eq.~\eqref{eq:sociswarmo}. As a result, opinions evolve slowly and may become trapped before reaching the usual bounded-confidence clustering.

\section{RADIUS OF THE CONSENSUS STATE}
\label{sec:consstate}

The results presented in the previous Section, have shown that in the consensus state, i.e., when all the agent share the same opinion, then they spatially organize in regular structure, i.e., a disk. The aim of this section is to provide an analytical expression for the radius of this circle as a function of the model parameters, $A$, $B$ and $J$. Let us observe that the radius will not depend on the parameters related to the opinion dynamics, $\lambda$ and $\mu$ because of the assumption of consensus, the only requirement will be for $d$ to be large enough to ensure the onset of this state.

Let us assume to have an infinite number of agents, i.e. $N\rightarrow \infty$, then $\rho(x,y,t)$ satisfies the following equations~\footnote{Let us observe that to lighten the notations, we omit the dependence on $O$ in $\rho$, because we will be interested in the full consensus case. To be more precise we should have written $\rho_{cons}(x,y)\delta(O-1/2)$, where $1/2$ is the consensus opinion because of the symmetry of the model.}
\begin{equation}
\label{eq:eqforrho}
\partial_t \rho + \nabla \cdot \left(\rho \vec{v}\right) = 0\text{ and } \int_{\mathbb{R}^2} \rho(x,y,t)dxdy = 1\quad \forall t\geq 0\, ,
\end{equation}
where the velocity field $\vec{v}$ is given by
\begin{widetext}
    \begin{equation}
\label{eq:velfield}
\vec{v}(x,y,t)=\int_{\mathbb{R}^2} \left[-(A+J)\frac{(x-x',y-y')}{\sqrt{(x-x')^2+(y-y')^2}}+B\frac{(x-x',y-y')}{(x-x')^2+(y-y')^2}\right)\rho(x',y',t)dx' dy'\, .
\end{equation}
\end{widetext}
Let us observe that because of the assumption $O_i=\hat{O}$, the exponential function multiplying $J$ in Eqs.~\eqref{eq:sociswarmx} and~\eqref{eq:sociswarmy}, simplifies to $1$. The leftmost  equation in~\eqref{eq:eqforrho} is the continuity equation, while the rightmost one encodes for the  preservation of the number of agents.

Let us denote by $\rho_{cons}(x,y)$ the stationary density at the consensus state, hence $\partial_t \rho_{cons}(x,y)=0$ and thus the velocity field at the consensus state must verify
    \begin{eqnarray}
\label{eq:divvelcons}
\nabla \cdot \left(\rho_{cons} \vec{v}_{cons}\right)&=&0 
\end{eqnarray}
Where
\begin{eqnarray}
\label{eq:velcons}
\vec{v}_{cons}(x,y)&=&\int_{\mathbb{R}^2} \left[-(A+J)\frac{(x-x',y-y')}{\sqrt{(x-x')^2+(y-y')^2}}+B\frac{(x-x',y-y')}{(x-x')^2+(y-y')^2}\right)\rho_{cons}(x',y')dx' dy'=0\, .
\end{eqnarray}

By using the property of the divergence operator $\nabla\cdot (f\vec{z})=\nabla f \cdot \vec{z}+f\nabla\cdot \vec{z}$, holding true for any smooth scalar function $f$ and a vector function $\vec{z}$, we hence obtain from Eq.~\eqref{eq:divvelcons} the following condition
\begin{equation}
\label{eq:velconssimp}
\nabla \cdot \vec{v}_{cons}=0\, ,
\end{equation}
where we used that $\vec{v}_{cons}=0$ and $\rho_{cons}> 0$.

Let us compute the divergence of $\vec{v}_{cons}$ given by~\eqref{eq:velcons}. It is composed by two terms, each one associated to one of the integrand functions. For the first one we have:
    \begin{eqnarray}
\label{eq:divv1}
\partial_x \frac{x-x'}{\sqrt{(x-x')^2+(y-y')^2}}+\partial_y \frac{y-y'}{\sqrt{(x-x')^2+(y-y')^2}}&=&
\frac{1}{\sqrt{(x-x')^2+(y-y')^2}}\, ,
\end{eqnarray}
while for the second one we can use the relation
\begin{equation}
\label{eq:divv2}
\nabla\cdot \frac{(x-x',y-y')}{(x-x')^2+(y-y')^2}=2\pi \delta_{x-x',y-y'}\, .
\end{equation}

In conclusion Eq.~\eqref{eq:velconssimp} rewrites
    \begin{equation}
\label{eq:velcons2}
0=-(A+J)\int_{\mathbb{R}^2} \frac{1}{\sqrt{(x-x')^2+(y-y')^2}}\rho_{cons}(x',y')dx' dy' +B\int_{\mathbb{R}^2}2\pi \delta_{x-x',y-y'}\rho_{cons}(x',y')dx' dy'\, ,
\end{equation}
namely
    \begin{equation}
\label{eq:velcons3}
\rho_{cons}(x,y)=\frac{A+J}{2\pi B}\int_{\mathbb{R}^2} \frac{1}{\sqrt{(x-x')^2+(y-y')^2}}\rho_{cons}(x',y')dx' dy' \, .
\end{equation}

In conclusion $\rho_{cons}(x,y)$ should be determined by using the latter equation, together with Eq.~\eqref{eq:velcons} and the mass preservation. To make some analytical progress, let us pass to polar coordinates, namely we introduce
\begin{equation}
\label{eq:polcoord}
x'=r\cos\theta \text{ and }y'=r\sin\theta \, ;
\end{equation}
moreover, the numerical simulations support the hypothesis that in the consensus state the distribution of the swarmalators is homogeneous with respect to the angle and with support in a disk of radius $R$, we can thus write (with a slight abuse of notation)
\begin{equation}
\label{eq:rhopol}
\rho_{cons}(r) = 
\begin{cases}
 \dfrac{\phi(r)}{2\pi} & \text{if $0\leq r \leq R$}\\
 0 & \text{if $r > R$}\, ,
\end{cases}
\end{equation}
for some unknown function $\phi(r)$. Let us observe that the factor $2\pi$ is the normalization with respect to the angles, namely from the mass conservation~\eqref{eq:eqforrho} we get
\begin{equation}
\label{eq:anglesnorm}
1=\int_{0}^{2\pi}d\theta \int_0^R  \frac{\phi(r)}{2\pi} r dr \Rightarrow 1= \int_0^R \phi(r) r dr\, .
\end{equation}
Let us also observe that homogeneity in the angular distribution allows to limit the computation of $\rho_{cons}(x,y)$ to the case $(x,y)=(x_1,0)$, $x_1>0$. We can thus rewrite~\eqref{eq:velcons3} as:
\begin{equation}
\label{eq:velcons4}
\rho_{cons}(x_1)=\frac{A+J}{2\pi B}\int_{0}^{2\pi}d\theta\int_0^R \frac{1}{\sqrt{x_1^2+r^2-2rx_1 \cos\theta}}\rho_{cons}(r)r dr \, .
\end{equation}

Let us now show how the integration over $\theta$ in the last integral returns an expression involving the complete elliptic integrals of the first kinds, given by
\begin{equation}
\label{eq:Kk}
K(\kappa):=\int_0^{\pi/2}\frac{1}{\sqrt{1-\kappa^2 \sin^2 t}}\, dt\, ,
\end{equation}
with $\kappa^2 <1$.

To prove this claim, let us consider the integral in Eq.~\eqref{eq:velcons4} involving $\theta$ and perform some algebraic operations
\begin{eqnarray}
\label{eq:Kk1}
I_1&=&\int_{0}^{2\pi}\frac{1}{\sqrt{x_1^2+r^2 - 2 r x_1 \cos\theta}}d\theta=\int_{0}^{2\pi}\frac{1}{\sqrt{x_1^2+r^2}}\frac{1}{\sqrt{1-\xi^2 \cos\theta}}d\theta\, ,
\end{eqnarray}
where we defined $\xi^2=\dfrac{2r x_1}{x_1^2+r^2}$. Let us observe that $\xi^2<1$, indeed
\begin{equation}
\label{eq:xi2less1}
\begin{aligned}
(x_1-r)^2>0
\Rightarrow x_1^2+r^2-2rx_1>0\Rightarrow x_1^2+r^2>2rx_1\Rightarrow 
\xi^2=\frac{2rx_1}{x_1^2+r^2}<1 .
\end{aligned}
\end{equation}
Let us now apply few trigonometric operations
\begin{eqnarray}
\label{eq:Kk2}
I_1 =\int_{-\pi}^{\pi}\frac{1}{\sqrt{x_1^2+r^2}}\frac{1}{\sqrt{1+\xi^2 \cos\theta}} d\theta =\int_{-\pi/2}^{\pi/2}\frac{2}{\sqrt{x_1^2+r^2}}\frac{1}{\sqrt{1+\xi^2 \cos(2\theta)}} d\theta =\int_{-\pi/2}^{\pi/2}\frac{2}{\sqrt{x_1^2+r^2}}\frac{1}{\sqrt{1+\xi^2 -2\xi^2\sin^2\theta}}d\theta\, .
\end{eqnarray}
By recalling the definition of $\xi$ we get
\begin{equation}
\label{eq:step1}
\frac{1}{\sqrt{x_1^2+r^2}}\frac{1}{\sqrt{1+\xi^2}}
=
\frac{1}{\sqrt{x_1^2+r^2}}
\frac{1}{\sqrt{1+\dfrac{2rx_1}{x_1^2+r^2}}}
=
\frac{1}{\sqrt{x_1^2+r^2}}
\frac{\sqrt{x_1^2+r^2}}{\sqrt{(x_1+r)^2}}
=
\frac{1}{\sqrt{(x_1+r)^2}}
=
\frac{1}{x_1+r}\, ,
\end{equation}
being $x_1+r>0$. Two last steps allow to write
\begin{eqnarray}
\label{eq:Kk3}
I_1 = 4\int_{0}^{\pi/2}\frac{1}{\sqrt{x_1^2+r^2}}\frac{1}{\sqrt{1+\xi^2 -2\xi^2\sin^2\theta}} d\theta =4\int_{0}^{\pi/2}\frac{1}{\sqrt{x_1^2+r^2}}\frac{1}{\sqrt{1+\xi^2}}\frac{1}{\sqrt{1 -2\frac{\xi^2}{1+\xi^2}\sin^2\theta}}d\theta\, .
\end{eqnarray}

 Let us finally define
\begin{equation}
\label{eq:kappa}
\kappa^2=2\frac{\xi^2}{1+\xi^2}=\frac{4rx_1}{(x_1+r)^2}\, ,
\end{equation}
and observe that $\xi^2<1$ implies that also $\kappa^2<1$. Eventually we obtain
\begin{equation}
\label{eq:Kk4}
I_1=\frac{4}{x_1+r}\int_{0}^{\pi/2}\frac{1}{\sqrt{1 -\kappa^2\sin^2\theta}}d\theta =\frac{4}{x_1+r}K(\kappa)\, .
\end{equation}

We have thus proved that Eq.~\eqref{eq:velcons4} is equivalent to
\begin{equation}
\label{eq:velcons5}
\rho_{cons}(x_1)=\frac{A+J}{2\pi B}\int_0^R \frac{4}{x_1+r}K(\kappa)\rho_{cons}(r)r dr \, ,
\end{equation}
where $\kappa(r)$ is given by~\eqref{eq:kappa}, or in terms of $\phi(r)$
\begin{equation}
\label{eq:velcons6}
\phi(x_1)=\frac{A+J}{2\pi B}\int_0^R \frac{4}{x_1+r}K(\kappa)\phi(r)r dr \, .
\end{equation}

An exact solution of the latter integral equation cannot be obtained, however we can determine a semi-analytical one by adapting a method inspired by the work by O'Keeffe \cite{o2017oscillators}. Eq.~\eqref{eq:velcons6} can be considered as an integral equation, for the unknown function $\phi$
\begin{equation}
\label{eq:velcons8}
\phi=\mathcal{M}\left(\phi\right) \, ,
\end{equation}
where, for all $s\in [0,R]$, $\mathcal{M}$ is the operator defined by
\begin{equation}
\label{eq:velcons9}
\mathcal{M}\left(\phi\right)(s):=\frac{A+J}{2\pi B}\int_0^R \frac{4r}{s+r}K\left(\sqrt{\frac{4rs}{(s+r)^2}}\right)\phi(r)dr \, .
\end{equation}
To remove the dependence on the model parameters, parameters $A$, $B$ and $J$, and on the radius $R$, let us define the auxiliary operator $\mathcal{M}^*$
\begin{equation}
\label{eq:velcons10}
\mathcal{M}^*\left(\phi\right)(s'):=\int_0^1 \frac{4r}{s'+r}K\left(\sqrt{\frac{4rs'}{(s'+r)^2}}\right)\phi(r)dr \quad \forall s'\in[0,1]\, ;
\end{equation}
it then easily follows that for $s'\in[0,1]$
    \begin{equation*}
\mathcal{M}\left(\phi\right)(s'R)=\frac{A+J}{2\pi B} \int_0^R \frac{4r}{s'R+r}K\left(\sqrt{\frac{4rs'R}{(s'R+r)^2}}\right)\phi(r)dr \\
=\frac{A+J}{2\pi B}R \int_0^1 \frac{4r'}{s'+r'}K\left(\sqrt{\frac{4r's'}{(s'+r')^2}}\right)\phi(r'R)dr'\, ,
\end{equation*}
hence
\begin{equation}
\label{eq:velcons11}
\mathcal{M}\left(\phi^*\right)=\frac{A+J}{2\pi B}R\mathcal{M}^*\left(\phi^*\right)\, ,
\end{equation}
where we introduced the notation $\phi^*(r')=\phi(r'R)$, for all $r'\in[0,1]$.

Stated differently, Eq.~\eqref{eq:velcons8} means that $\phi$ is an eigenfunction of $\mathcal{M}$ associated to the eigenvalue $1$; hence - thanks to the latter relation - if $\phi^*$ is an eigenfunction of $\mathcal{M}^*$ with eigenvalue $\beta$, i.e., $\mathcal{M}^*\phi^*(s')=\beta \phi^*(s')$, for all $s'\in[0,1]$, then by choosing $R$ such that
\begin{equation}
\label{eq:Rchoice}
R\frac{A+J}{2\pi B}\beta = 1\, ,
\end{equation}
namely
\begin{equation}
\label{eq:Rchoice2}
R = 2\pi\frac{B}{\beta(A+J)}\, ,
\end{equation}
we have found an eigenfunction $\phi$ of $\mathcal{M}$ with eigenvalue $1$. 

Because the operator $\mathcal{M}^*$ depends only on the complete elliptic integral of the first kind, and not on the model parameters, $A$, $B$, $J$ nor $R$, then $\beta$ does the same, it is thus an ``universal'' constant. We can thus conclude that~\eqref{eq:Rchoice2} returns the analytical form of the  radius of the consensus domain as a function of the model parameter.

To determine $\beta$ we compute the integral~\eqref{eq:velcons10} by discretizing the interval $[0,1]$, i.e., let $n\gg 1$, $i=1,\dots,n$ and $r_i=(i-1)/(n-1)$, hence for all $j\in\{1,\dots,n\}$
\begin{equation}
\label{eq:velcons6dis}
\phi_j=\frac{1}{n}\sum_{i=1}^{n} \frac{4r_i}{r_j+r_i}K\left(\frac{4r_ir_j}{(r_i+r_j)^2}\right)\phi_i \, ,
\end{equation}
where we denoted $\phi_i=\phi(r_i)$, let us observe that the sum does not contain the index $i=j$, indeed in this case $\kappa=1$ and the complete elliptic integral diverges. 
In conclusion we get
\begin{equation}
\label{eq:velcons6dis2}
\phi_j=\sum_{i=1}^{n} M^*_{ji}\phi_i \, ,
\end{equation}
where we have defined the matrix $\mathbf{M}^*$ by
\begin{equation}
\label{eq:matrixM}
M^*_{ji}:=\frac{1}{n}\frac{4r_i }{r_j+r_i}K\left(\sqrt{\frac{4r_ir_j}{(r_i+r_j)^2}}\right)\, .
\end{equation}

\begin{figure}[ht!]
        \includegraphics[width=0.4\textwidth]{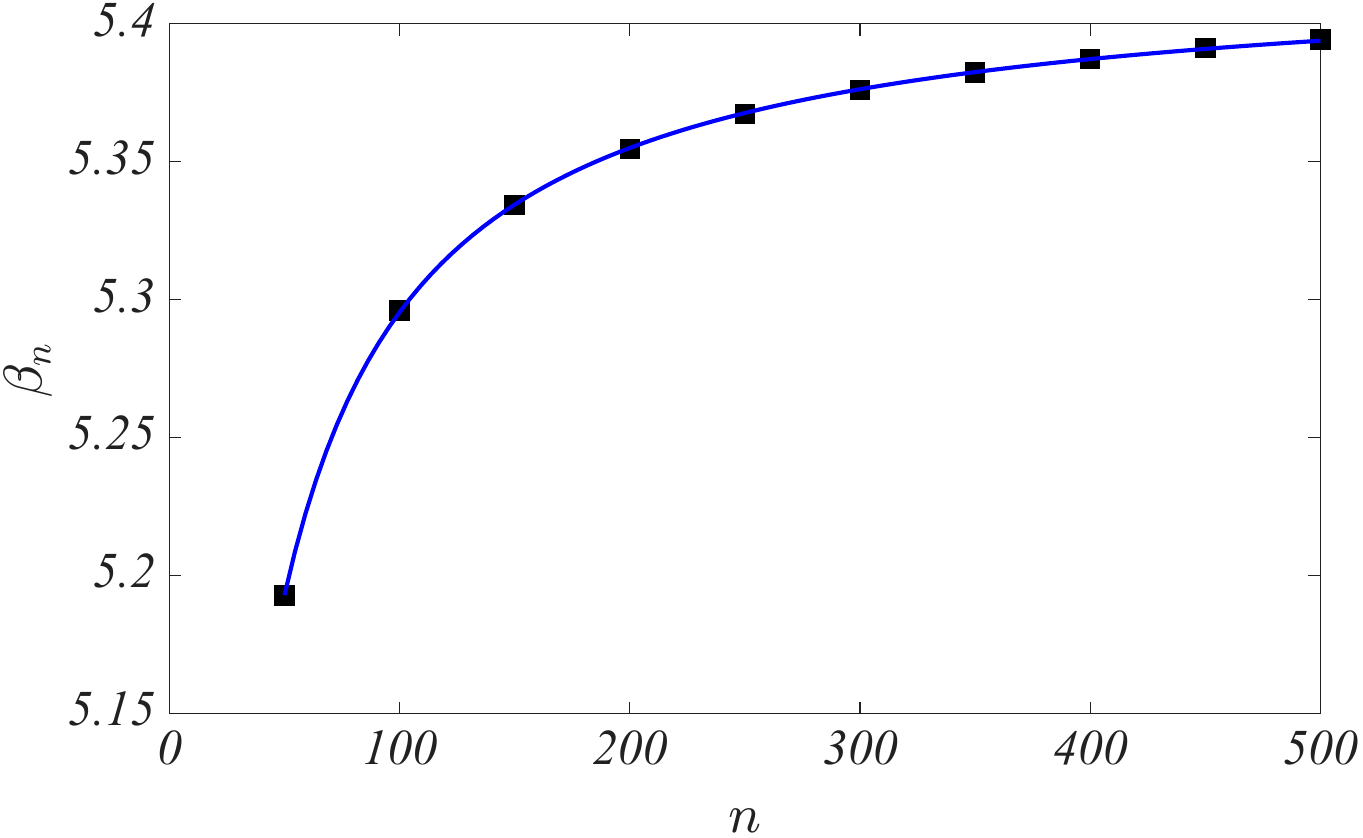}
        \caption{{\bf Evolution of largest eigenvalue of $\mathbf{M}^*$}. We display the dependence of the largest eigenvalue $\beta_n$ of $\mathbf{M}^*$ as a function of the discretization mesh of the integral, $n$. Black squares are the numerically computed values of $\beta_n$ by assuming a mesh size $n$, the blue curve is the numerical fit $a-b/(c+n)$ with $a=5.4213$, $b = 13.9992$ and $c = 11.2961$.}
    \label{fig:betavsn}
\end{figure}

In Fig.~\ref{fig:betavsn} we report the dependence of the largest eigenvalue $\beta_n$ of $\mathbf{M}^*$ as a function of the mesh size used to discretize the integral. We also display the numerical fit that shows a very good approximation with the function $a-b/(c+n)$ with $a = 5.4213$, $b = 13.9992$ and $c = 11.2961$. We can thus conclude that in the limit $n\rightarrow \infty$, the largest eigenvalue $\beta$ of the operator $\mathcal{M}^*$ will assume the value $\beta_n\rightarrow \beta=5.4213$.

\begin{figure}[ht!]
        \includegraphics[width=0.4\textwidth]{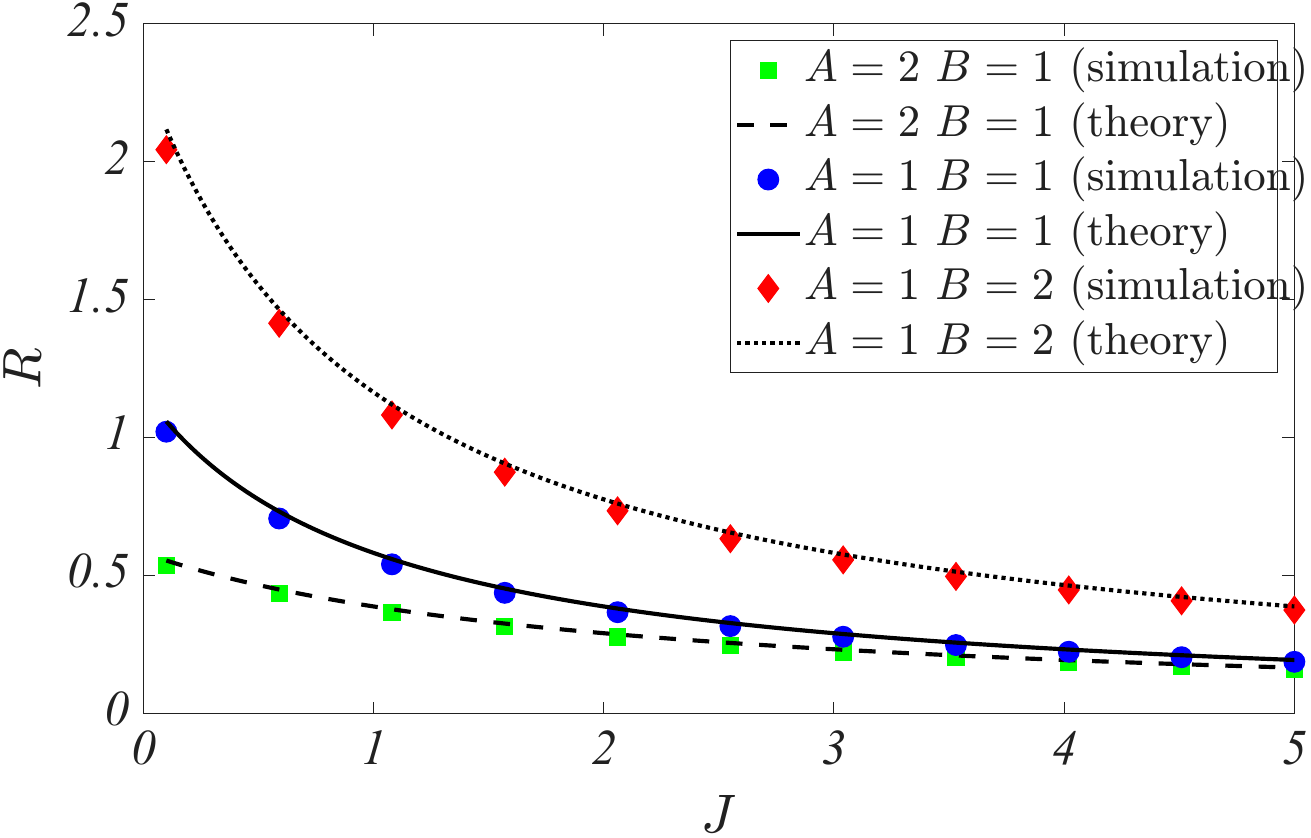}
        \caption{{\bf Radius of the disk of consensus as a function of $J$}. We compare the analytical formula for the radius of the consensus state with the same quantity obtained from numerical simulations  for three sets of parameters $(A,B)=(1,1)$ (blue circle), $(A,B)=(2,1)$ (green square) and $(A,B)=(1,2)$ (red diamond), the curves represent the theory presented above for the same values of the parameters.}
    \label{fig:Rthexp}
\end{figure}

By using this asymptotic value $\beta=5.4213$, we can compute the radius $R$ from~\eqref{eq:Rchoice2} as a function of $J$ for three sets of parameters $(A,B)=(1,1)$, $(A,B)=(2,1)$ and $(A,B)=(1,2)$; the results reported in Fig.~\ref{fig:Rthexp} show an excellent agreement between the theory and the numerical results.

\section{Conclusion}
\label{sec:conclusion}
In this work, we introduced a social swarm model that couples spatial aggregation with bounded-confidence opinion dynamics. Instead of describing the internal state by a Kuramoto oscillator, we used a continuous-time version of the Deffuant model, in which agents  adjust their opinions only when their opinion difference is below a confidence threshold. This modification gives the internal variable a direct social interpretation while preserving the mutual feedback between internal dynamics and spatial motion. In this way the proposed model is more suitable to describe systems where agents exhibit social interactions, such as opinion exchange.

The numerical results show that the confidence bound $d$ controls the number of opinion groups formed by the population. When $d$ is small enough, several opinion clusters can appear, consistently with the bounded-confidence mechanism. The parameter $\lambda$ controls how strongly opinion differences affect spatial attraction. Small values of $\lambda$ allow different opinion groups to remain close in space, whereas large values promote spatial separation between groups with different opinions.  The parameter $J$ also affects the spatial distribution of the groups. When $J>0$, agents are more attracted to each other, and the opinion groups tend to be more separated in space, especially for large $\lambda$. When $J<0$, the attraction between agents can become weak. In this case, the groups can overlap more, the opinions can change more slowly, and the number of opinion groups can become larger, especially for small $\lambda$.


The numerical results suggest that once the social swarm reach a consensus state they spatially organize into a disk with radius $R$. We have been able to obtain an analytical expression for the radius $R$ when considering a nonlinear attraction kernel, and we found that, $R=\frac{2\pi B}{\beta (A+J)}$ where $\beta$ is obtained from a rescaled eigenvalue problem. The theoretical prediction agrees well with numerical simulations for different values of $A$, $B$, and $J$.

Overall, this work shows that the proposed model provides a flexible framework for studying the interaction between social compromise and spatial self-organization. In particular, it makes it possible to control both the number of opinion groups and their spatial separation. Future work could investigate heterogeneous confidence bounds and other opinion models, as well as how bidirectional coupling between swarming dynamics and opinion dynamics affects the collective dynamics.

\bibliographystyle{unsrt}
\bibliography{sample}

@article{deffuant2000mixing,
  title={Mixing beliefs among interacting agents},
  author={Deffuant, Guillaume and Neau, David and Amblard, Frederic and Weisbuch, G{\'e}rard},
  journal={Advances in Complex Systems},
  volume={3},
  number={01n04},
  pages={87--98},
  year={2000},
  publisher={World Scientific}
}

@article{GPKG2024,
title = {Amplitude responses of swarmalators},
author = {Ghosh, Samali and Pal,  Suvam and Kumar Sar, Gourab and Ghosh, Dibakar},
  journal={Physical Review E},
  volume={109},
  pages={054205},
  year={2024},
  publisher={APS}
  }

@article{o2022collective,
  title={Collective behavior of swarmalators on a ring},
  author={O'Keeffe, Kevin and Ceron, Steven and Petersen, Kirstin},
  journal={Physical Review E},
  volume={105},
  number={1},
  pages={014211},
  year={2022},
  publisher={APS}
}

@article{o2026time,
  title={Time delay in the 1d swarmalator model},
  author={O'Keeffe, KP and Hindes, Jason},
  journal={arXiv preprint arXiv:2602.08156},
  year={2026}
}

@article{o2025global,
  title={Global synchronization theorem for coupled swarmalators},
  author={O’Keeffe, Kevin},
  journal={Chaos: An Interdisciplinary Journal of Nonlinear Science},
  volume={35},
  number={2},
  pages={023150},
  year={2025},
  publisher={AIP Publishing}
}

@article{o2025stability,
  title={Stability of the 1D swarmalator model in the continuum limit},
  author={O’Keeffe, Kevin},
  journal={Chaos: An Interdisciplinary Journal of Nonlinear Science},
  volume={35},
  number={7},
  pages={073139},
  year={2025},
  publisher={AIP Publishing}
}

@article{o2017oscillators,
  title={Oscillators that sync and swarm},
  author={O’Keeffe, Kevin P and Hong, Hyunsuk and Strogatz, Steven H},
  journal={Nature Communications},
  volume={8},
  number={1},
  pages={1504},
  year={2017},
  publisher={Nature Publishing Group UK London}
}

@article{sar2026interplay,
  title={Interplay of sync and swarm: Theory and application of swarmalators},
  author={Sar, Gourab Kumar and O’Keeffe, Kevin and Liz{\'a}rraga, Joao UF and de Aguiar, Marcus AM and Bettstetter, Christian and Ghosh, Dibakar},
  journal={Physics Reports},
  volume={1167},
  pages={1--52},
  year={2026},
  publisher={Elsevier}
}

@article{blum2024swarmalators,
  title={Swarmalators with delayed interactions},
  author={Blum, Nicholas and Li, Andre and O'Keeffe, Kevin and Kogan, Oleg},
  journal={Physical Review E},
  volume={109},
  number={1},
  pages={014205},
  year={2024},
  publisher={APS}
}

@article{o2022swarmalators,
  title={Swarmalators on a ring with distributed couplings},
  author={O'Keeffe, Kevin and Hong, Hyunsuk},
  journal={Physical Review E},
  volume={105},
  number={6},
  pages={064208},
  year={2022},
  publisher={APS}
}

@article{hao2023attractive,
  title={Attractive and repulsive interactions in the one-dimensional swarmalator model},
  author={Hao, Baoli and Zhong, Ming and O'Keeffe, Kevin},
  journal={Physical Review E},
  volume={108},
  number={6},
  pages={064214},
  year={2023},
  publisher={APS}
}

@article{sar2023pinning,
  title={Pinning in a system of swarmalators},
  author={Sar, Gourab Kumar and Ghosh, Dibakar and O'Keeffe, Kevin},
  journal={Physical Review E},
  volume={107},
  number={2},
  pages={024215},
  year={2023},
  publisher={APS}
}

@article{sar2023swarmalators,
  title={Swarmalators on a ring with uncorrelated pinning},
  author={Sar, Gourab Kumar and O’Keeffe, Kevin and Ghosh, Dibakar},
  journal={Chaos: An Interdisciplinary Journal of Nonlinear Science},
  volume={33},
  number={11},
  pages={111103},
  year={2023},
  publisher={AIP Publishing}
}

@article{anwar2024forced,
  title={Forced one-dimensional swarmalator model},
  author={Anwar, Md Sayeed and Ghosh, Dibakar and O'Keeffe, Kevin},
  journal={Physical Review E},
  volume={110},
  number={5},
  pages={054205},
  year={2024},
  publisher={APS}
}

@article{anwar2025forced,
  title={Forced swarmalators that move in higher-dimensional spaces},
  author={Anwar, Md Sayeed and Ghosh, Dibakar and O'Keeffe, Kevin},
  journal={Physical Review E},
  volume={111},
  number={4},
  pages={044207},
  year={2025},
  publisher={APS}
}

@article{lizarraga2023synchronization,
  title={Synchronization of Sakaguchi swarmalators},
  author={Liz{\'a}rraga, Joao UF and de Aguiar, Marcus AM},
  journal={Physical Review E},
  volume={108},
  number={2},
  pages={024212},
  year={2023},
  publisher={APS}
}

@article{sar2025effects,
  title={Effects of coupling range on the dynamics of swarmalators},
  author={Sar, Gourab Kumar and O'Keeffe, Kevin and Ghosh, Dibakar},
  journal={Physical Review E},
  volume={111},
  number={2},
  pages={024206},
  year={2025},
  publisher={APS}
}

@article{hong2023swarmalators,
  title={Swarmalators with thermal noise},
  author={Hong, Hyunsuk and O'Keeffe, Kevin P and Lee, Jae Sung and Park, Hyunggyu},
  journal={Physical Review Research},
  volume={5},
  number={2},
  pages={023105},
  year={2023},
  publisher={APS}
}

@article{lr2r-ynzs,
  title = {Dynamics of pulsating swarmalators on a ring},
  author = {Ghosh, Samali and O'Keeffe, Kevin and Sar, Gourab Kumar and Ghosh, Dibakar},
  journal = {Phys. Rev. E},
  volume = {112},
  issue = {5},
  pages = {054217},
  numpages = {11},
  year = {2025},
  month = {Nov},
  publisher = {American Physical Society}
}

@article{anwar2024collective,
  title={Collective dynamics of swarmalators with higher-order interactions},
  author={Anwar, Md Sayeed and Sar, Gourab Kumar and Perc, Matja{\v{z}} and Ghosh, Dibakar},
  journal={Communications Physics},
  volume={7},
  number={1},
  pages={59},
  year={2024},
  publisher={Nature Publishing Group UK London}
}

@article{senthamizhan2026swarmalators,
  title={Swarmalators with frequency-weighted interactions},
  author={Senthamizhan, R and Gopal, R and Chandrasekar, VK},
  journal={Physical Review E},
  volume={113},
  number={3},
  pages={034216},
  year={2026},
  publisher={APS}
}

@article{sar2025strategy,
  title={Strategy to control synchronized dynamics in swarmalator systems},
  author={Sar, Gourab Kumar and Anwar, Md Sayeed and Moriam{\'e}, Martin and Ghosh, Dibakar and Carletti, Timoteo},
  journal={Physical Review E},
  volume={111},
  number={3},
  pages={034212},
  year={2025},
  publisher={APS}
}

@article{yoon2022sync,
  title={Sync and swarm: Solvable model of nonidentical swarmalators},
  author={Yoon, S and O’Keeffe, KP and Mendes, JFF and Goltsev, AV},
  journal={Physical Review Letters},
  volume={129},
  number={20},
  pages={208002},
  year={2022},
  publisher={APS}
}

@article{o2026unsteady,
  title={Unsteady phase waves in the 1D swarmalator model with inertia},
  author={O'Keeffe, Kevin P},
  journal={arXiv preprint arXiv:2603.12531},
  year={2026}
}

@article{ghosh2026emergent,
  title={Emergent dynamics in heterogeneous pulsatile swarmalators},
  author={Ghosh, Samali and O’Keeffe, Kevin and Ghosh, Dibakar},
  journal={Chaos: An Interdisciplinary Journal of Nonlinear Science},
  volume={36},
  number={3},
  year={2026},
  publisher={AIP Publishing}
}

@article{lambu2026delay,
  title={Delay induced double explosive transition in a swarmalator system},
  author={Lambu, Carmel T and Mbonwouo, Romuald T and Simo, Gael R and Jiofack, Delors A and Kongni, Steve J and Louodop, Patrick and Cerdeira, Hilda A},
  journal={Physical Review E},
  volume={113},
  number={2},
  pages={024203},
  year={2026},
  publisher={APS}
}

@article{kongni2023phase,
  title={Phase transitions on a multiplex of swarmalators},
  author={Kongni, Steve J and Nguefoue, Venceslas and Njougouo, Thierry and Louodop, Patrick and Ferreira, Fernando Fagundes and Tchitnga, Robert and Cerdeira, Hilda A},
  journal={Physical Review E},
  volume={108},
  number={3},
  pages={034303},
  year={2023},
  publisher={APS}
}

@article{kongni2024expected,
  title={Expected and unexpected routes to synchronization in a system of swarmalators},
  author={Kongni, Steve J and Njougouo, Thierry and Louodop, Patrick and Tchitnga, Robert and Ferreira, Fernando F and Cerdeira, Hilda A},
  journal={Physical Review E},
  volume={110},
  number={6},
  pages={L062301},
  year={2024},
  publisher={APS}
}

@article{smith2024swarmalators,
  title={Swarmalators with higher harmonic coupling: Clustering and vacillating},
  author={Smith, Lauren D},
  journal={SIAM Journal on Applied Dynamical Systems},
  volume={23},
  number={2},
  pages={1133--1158},
  year={2024},
  publisher={SIAM}
}

@article{Bialek2012,
  author  = {Bialek, William and Cavagna, Andrea and Giardina, Irene and Mora, Thierry and Silvestri, Edmondo and Viale, Massimiliano and Walczak, Aleksandra M.},
  title   = {Statistical mechanics for natural flocks of birds},
  journal = {Proceedings of the National Academy of Sciences},
  volume  = {109},
  pages   = {4786--4791},
  year    = {2012}
}

@article{Hemelrijk2012,
  author  = {Hemelrijk, Charlotte K. and Hildenbrandt, Hanno},
  title   = {Schools of fish and flocks of birds: their shape and internal structure by self-organization},
  journal = {Interface Focus},
  volume  = {2},
  pages   = {726--737},
  year    = {2012}
}

@book{Sumpter2010,
  author    = {Sumpter, David J. T.},
  title     = {Collective Animal Behavior},
  publisher = {Princeton University Press},
  year      = {2010}
}

@article{Okubo1986,
  author  = {Okubo, Akira},
  title   = {Dynamical aspects of animal grouping: swarms, schools, flocks, and herds},
  journal = {Advances in Biophysics},
  volume  = {22},
  pages   = {1--94},
  year    = {1986}
}

@article{Norris1988,
  author  = {Norris, Kenneth S. and Schilt, Carl R.},
  title   = {Cooperative societies in three-dimensional space: on the origins of aggregations, flocks, and schools, with special reference to dolphins and fish},
  journal = {Ethology and Sociobiology},
  volume  = {9},
  pages   = {149--179},
  year    = {1988}
}

@article{Toner1998,
  author  = {Toner, John and Tu, Yuhai},
  title   = {Flocks, herds, and schools: A quantitative theory of flocking},
  journal = {Physical Review E},
  volume  = {58},
  pages   = {4828},
  year    = {1998}
}

@article{Toner2005,
  author  = {Toner, John and Tu, Yuhai and Ramaswamy, Sriram},
  title   = {Hydrodynamics and phases of flocks},
  journal = {Annals of Physics},
  volume  = {318},
  pages   = {170--244},
  year    = {2005}
}

@article{Cucker2007,
  author  = {Cucker, Felipe and Smale, Steve},
  title   = {Emergent behavior in flocks},
  journal = {IEEE Transactions on Automatic Control},
  volume  = {52},
  pages   = {852--862},
  year    = {2007}
}

@article{Vicsek2012,
  author  = {Vicsek, Tam{\'a}s and Zafeiris, Anna},
  title   = {Collective motion},
  journal = {Physics Reports},
  volume  = {517},
  pages   = {71--140},
  year    = {2012}
}

@book{Miller2010,
  author    = {Miller, Peter},
  title     = {The Smart Swarm: How Understanding Flocks, Schools, and Colonies Can Make Us Better at Communicating, Decision Making, and Getting Things Done},
  publisher = {Avery Publishing Group},
  year      = {2010}
}

@article{VanDerHoek2008,
  author  = {Van der Hoek, Wiebe and Wooldridge, Michael},
  title   = {Multi-agent systems},
  journal = {Foundations of Artificial Intelligence},
  volume  = {3},
  pages   = {887--928},
  year    = {2008}
}

@article{Brambilla2013,
  author  = {Brambilla, Manuele and Ferrante, Eliseo and Birattari, Mauro and Dorigo, Marco},
  title   = {Swarm robotics: a review from the swarm engineering perspective},
  journal = {Swarm Intelligence},
  volume  = {7},
  pages   = {1--41},
  year    = {2013}
}

@incollection{Kennedy2006,
  author    = {Kennedy, James},
  title     = {Swarm intelligence},
  booktitle = {Handbook of Nature-Inspired and Innovative Computing},
  publisher = {Springer},
  pages     = {187--219},
  year      = {2006}
}

@inproceedings{Reynolds1987,
  author    = {Reynolds, Craig W.},
  title     = {Flocks, herds and schools: A distributed behavioral model},
  booktitle = {Proceedings of the 14th Annual Conference on Computer Graphics and Interactive Techniques},
  pages     = {25--34},
  year      = {1987}
}

@article{Hartman2006,
  author  = {Hartman, Christopher and Benes, Bedrich},
  title   = {Autonomous boids},
  journal = {Computer Animation and Virtual Worlds},
  volume  = {17},
  pages   = {199--206},
  year    = {2006}
}

@article{Vicsek1995,
  author  = {Vicsek, Tam{\'a}s and Czir{\'o}k, Andr{\'a}s and Ben-Jacob, Eshel and Cohen, Inon and Shochet, Ofer},
  title   = {Novel type of phase transition in a system of self-driven particles},
  journal = {Physical Review Letters},
  volume  = {75},
  pages   = {1226},
  year    = {1995}
}

@article{Carrillo2010,
  author  = {Carrillo, Jos{\'e} A. and Fornasier, Massimo and Toscani, Giuseppe and Vecil, Francesco},
  title   = {Particle, kinetic, and hydrodynamic models of swarming},
  journal = {Mathematical Modeling of Collective Behavior in Socio-economic and Life Sciences},
  pages   = {297--336},
  year    = {2010}
}

@article{Winfree1967,
  author  = {Winfree, Arthur T.},
  title   = {Biological rhythms and the behavior of populations of coupled oscillators},
  journal = {Journal of Theoretical Biology},
  volume  = {16},
  pages   = {15--42},
  year    = {1967}
}

@inproceedings{Kuramoto1975,
  author    = {Kuramoto, Yoshiki},
  title     = {Self-entrainment of a population of coupled non-linear oscillators},
  booktitle = {International Symposium on Mathematical Problems in Theoretical Physics},
  publisher = {Springer},
  pages     = {420--422},
  year      = {1975}
}

@book{Pikovsky2001,
  author    = {Pikovsky, Arkady and Rosenblum, Michael and Kurths, J{\"u}rgen},
  title     = {Synchronization: A Universal Concept in Nonlinear Sciences},
  publisher = {Cambridge University Press},
  year      = {2001}
}

@article{Buck1988,
  author  = {Buck, John},
  title   = {Synchronous rhythmic flashing of fireflies. II},
  journal = {The Quarterly Review of Biology},
  volume  = {63},
  pages   = {265--289},
  year    = {1988}
}

@article{Neda2000,
  author  = {N{\'e}da, Zolt{\'a}n and Ravasz, Erzs{\'e}bet and Brechet, Yves and Vicsek, Tam{\'a}s and Barab{\'a}si, Albert-L{\'a}szl{\'o}},
  title   = {The sound of many hands clapping},
  journal = {Nature},
  volume  = {403},
  pages   = {849--850},
  year    = {2000}
}

@article{Womelsdorf2007,
  author  = {Womelsdorf, Thilo and Fries, Pascal},
  title   = {The role of neuronal synchronization in selective attention},
  journal = {Current Opinion in Neurobiology},
  volume  = {17},
  pages   = {154--160},
  year    = {2007}
}

@article{Rohden2012,
  author  = {Rohden, Martin and Sorge, Andreas and Timme, Marc and Witthaut, Dirk},
  title   = {Self-organized synchronization in decentralized power grids},
  journal = {Physical Review Letters},
  volume  = {109},
  pages   = {064101},
  year    = {2012}
}

@article{Bernoff2013,
  author  = {Bernoff, Andrew J. and Topaz, Chad M.},
  title   = {Nonlocal aggregation models: A primer of swarm equilibria},
  journal = {SIAM Review},
  volume  = {55},
  pages   = {709--747},
  year    = {2013}
}

@article{Topaz2006,
  author  = {Topaz, Chad M. and Bertozzi, Andrea L. and Lewis, Mark A.},
  title   = {A nonlocal continuum model for biological aggregation},
  journal = {Bulletin of Mathematical Biology},
  volume  = {68},
  pages   = {1601--1623},
  year    = {2006}
}

@article{Topaz2004,
  author  = {Topaz, Chad M. and Bertozzi, Andrea L.},
  title   = {Swarming patterns in a two-dimensional kinematic model for biological groups},
  journal = {SIAM Journal on Applied Mathematics},
  volume  = {65},
  pages   = {152--174},
  year    = {2004}
}

@article{Frasca2008,
  author  = {Frasca, Mattia and Buscarino, Arturo and Rizzo, Alessandro and Fortuna, Luigi and Boccaletti, Stefano},
  title   = {Synchronization of moving chaotic agents},
  journal = {Physical Review Letters},
  volume  = {100},
  pages   = {044102},
  year    = {2008}
}

@article{Chowdhury2019,
  author  = {Chowdhury, S. N. and Majhi, S. and Ozer, M. and Ghosh, Dibakar and Perc, Matja{\v{z}}},
  title   = {Synchronization to extreme events in moving agents},
  journal = {New Journal of Physics},
  volume  = {21},
  pages   = {073048},
  year    = {2019}
}

@article{Majhi2019,
  author  = {Majhi, Soumen and Ghosh, Dibakar and Kurths, J{\"u}rgen},
  title   = {Emergence of synchronization in multiplex networks of mobile R{\"o}ssler oscillators},
  journal = {Physical Review E},
  volume  = {99},
  pages   = {012308},
  year    = {2019}
}

@article{Sar2022Competitive,
  author  = {Sar, Gourab Kumar and Chowdhury, S. N. and Perc, Matja{\v{z}} and Ghosh, Dibakar},
  title   = {Swarmalators under competitive time-varying phase interactions},
  journal = {New Journal of Physics},
  volume  = {24},
  pages   = {043004},
  year    = {2022}
}

@article{djeudjo2026role,
  title={The role of asymmetric time delay and its structure in 1D swarmalators},
  author={Djeudjo, Rommel Tchinda and Sar, Gourab Kumar and Carletti, Timoteo},
  journal={arXiv preprint arXiv:2605.11713},
  year={2026}
}

@article{sar2022dynamics,
  title={Dynamics of swarmalators: A pedagogical review},
  author={Sar, Gourab Kumar and Ghosh, Dibakar},
  journal={Europhysics Letters},
  volume={139},
  number={5},
  pages={53001},
  year={2022},
  publisher={EDP Sciences, IOP Publishing and Societ{\`a} Italiana di Fisica}
}

\appendix
\onecolumngrid

\setcounter{equation}{0}
\renewcommand{\theequation}{A\arabic{equation}}
\setcounter{figure}{0}
\renewcommand{\thefigure}{A\arabic{figure}}

\section{Case of the linear attraction kernel}
\label{app:linear_kernel}

This appendix collects complementary results for a variant of the social swarmalator model in which the attraction term is linear. The purpose is not to repeat the analysis of the main text, but to show the robustness to a change in the spatial kernel and which behaviors are specific to the linear case. Unless otherwise stated, we use the same notation and parameter conventions as in the main text, with $A=B=1$.

\subsection{The model}
\label{app:linear_kernel_model}

Let $\mathbf r_i=(x_i,y_i)\in\mathbb R^2$, $\mathbf r_{ij}=\mathbf r_j-\mathbf r_i$, $r_{ij}=\|\mathbf r_{ij}\|$, and $O_{ij}=O_j-O_i$. Compared to Eqs.~\eqref{eq:sociswarmx}--\eqref{eq:sociswarmo}, only the spatial attraction kernel is modified, the unit vector $\mathbf r_{ij}/r_{ij}$ is replaced by the displacement vector $\mathbf r_{ij}$. The model is therefore given by
\begin{align}
\dot{\mathbf r}_i
&=\frac{1}{N}\sum_{j\neq i}
\left[
\mathbf r_{ij}\left(A+J e^{-\lambda |O_{ij}|}\right)
-B\frac{\mathbf r_{ij}}{r_{ij}^{2}}
\right],
\label{eq:linear_kernel_spatial}\\
\dot O_i
&=\frac{\mu}{N}\sum_{j\neq i}
\frac{\Theta\left(d-|O_{ij}|\right)O_{ij}}{r_{ij}}.
\label{eq:linear_kernel_opinion}
\end{align}
The parameter $d$ still defines the bounded-confidence threshold in opinion space, whereas $\lambda$ controls how strongly opinion differences reduce the attractive spatial coupling.

\subsection{Numerical behavior}
\label{app:linear_kernel_numerics}

\begin{figure*}[ht!]
    \centering
    \begin{tabular}{ccccc}
        \includegraphics[width=0.19\textwidth]{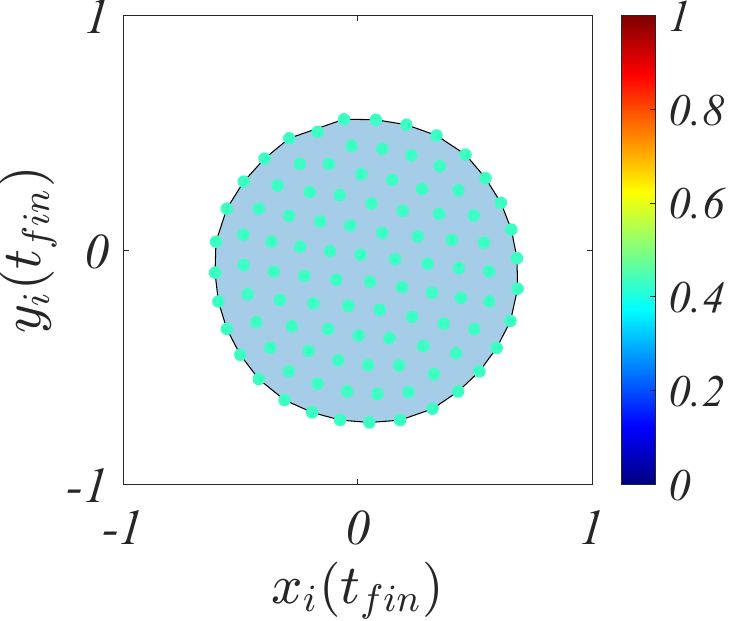} &
        \includegraphics[width=0.19\textwidth]{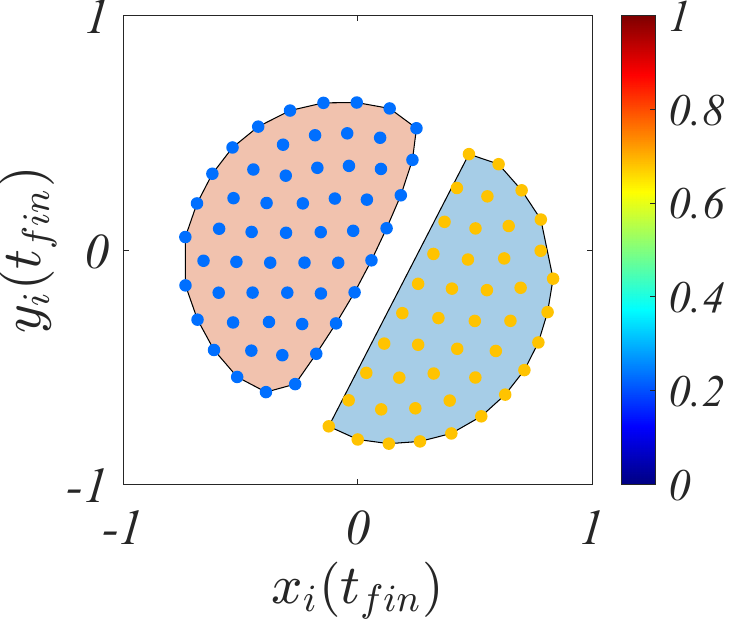} &
        \includegraphics[width=0.19\textwidth]{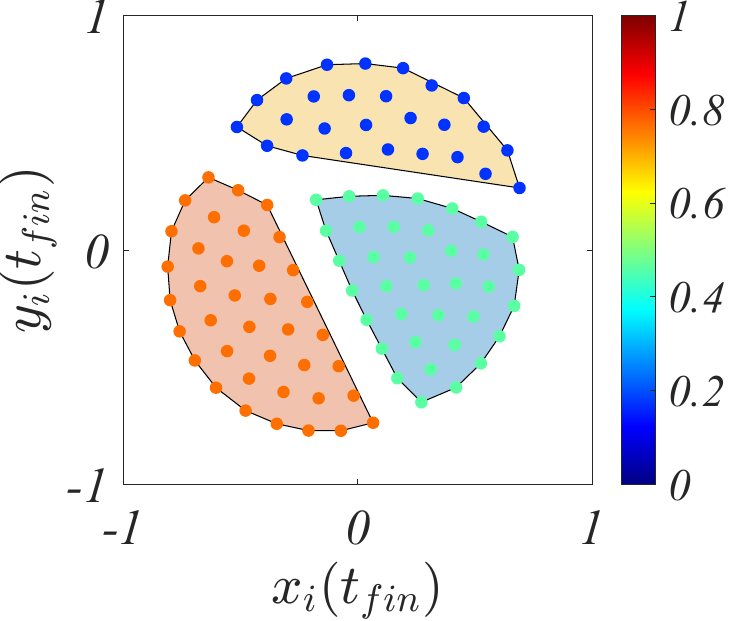} &
        \includegraphics[width=0.19\textwidth]{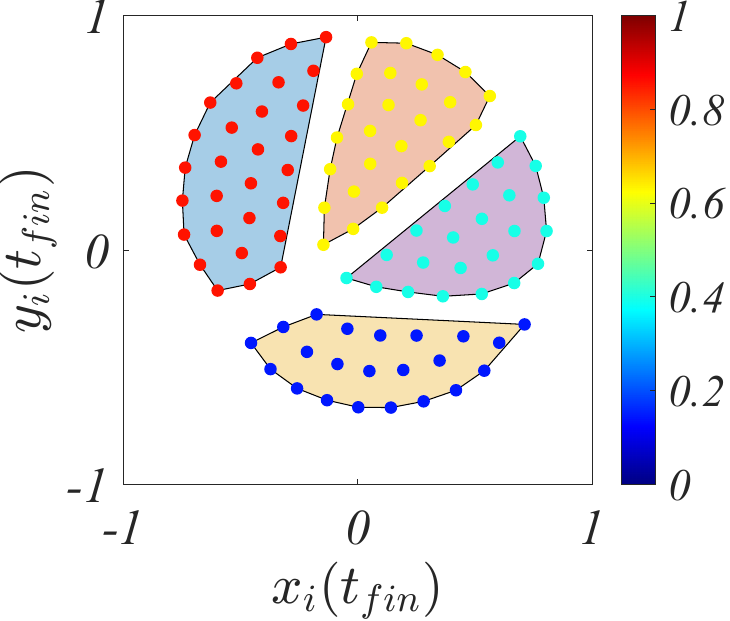} &
        \includegraphics[width=0.19\textwidth]{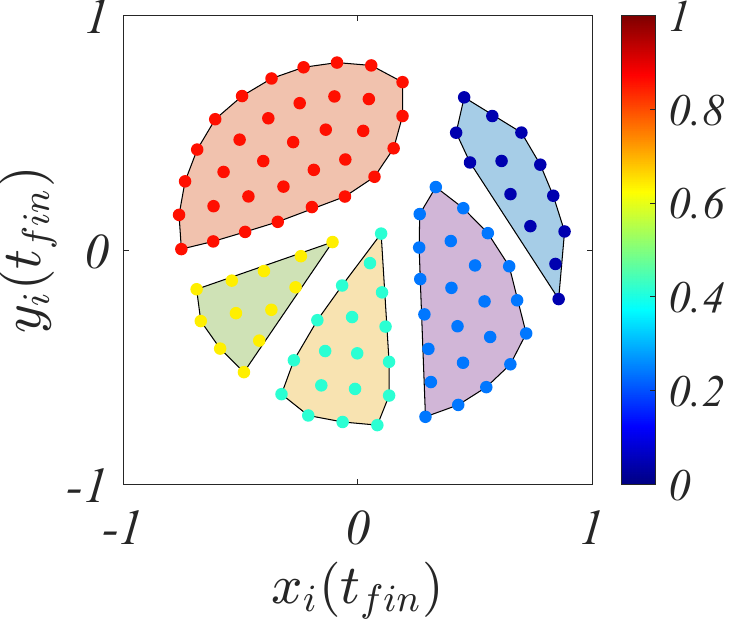}\\
        \textbf{(a1)} & \textbf{(b1)} & \textbf{(c1)} & \textbf{(d1)} & \textbf{(e1)} \\

        \includegraphics[width=0.19\textwidth]{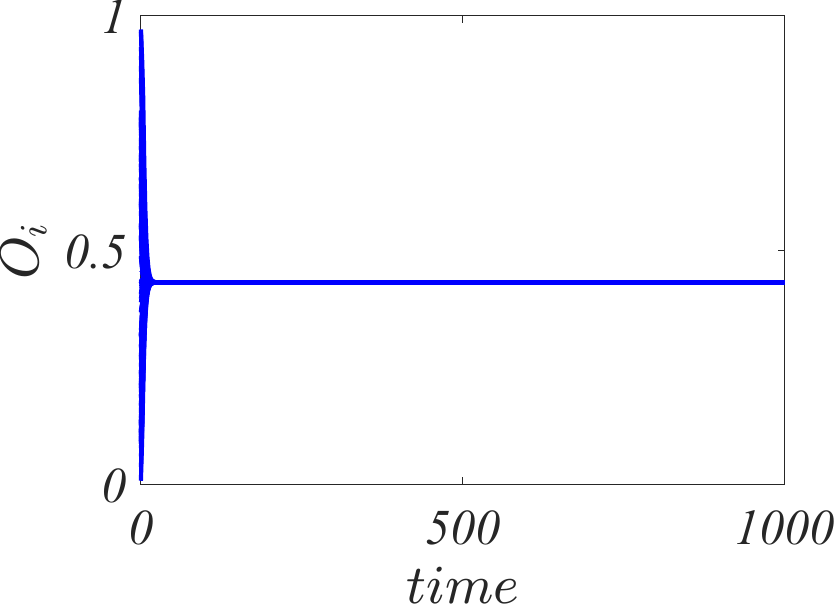} &
        \includegraphics[width=0.19\textwidth]{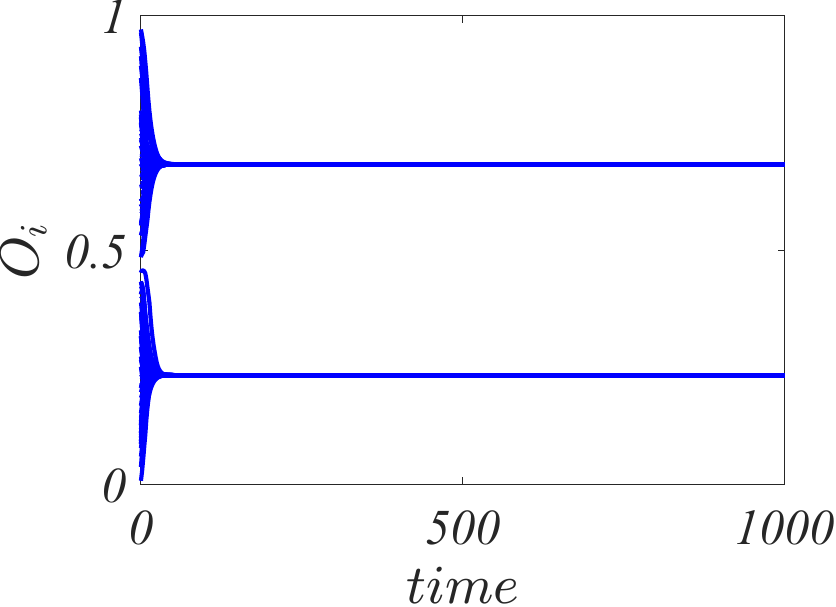} &
        \includegraphics[width=0.19\textwidth]{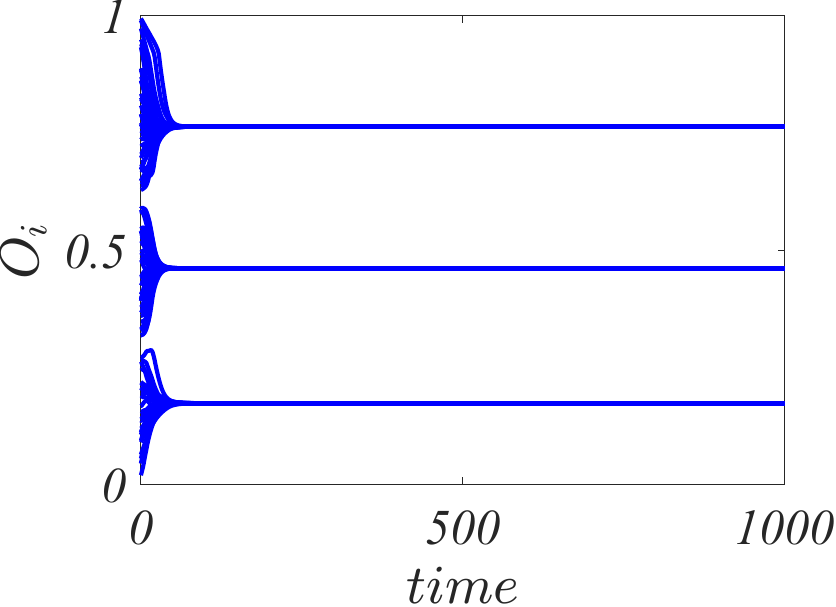} &
        \includegraphics[width=0.19\textwidth]{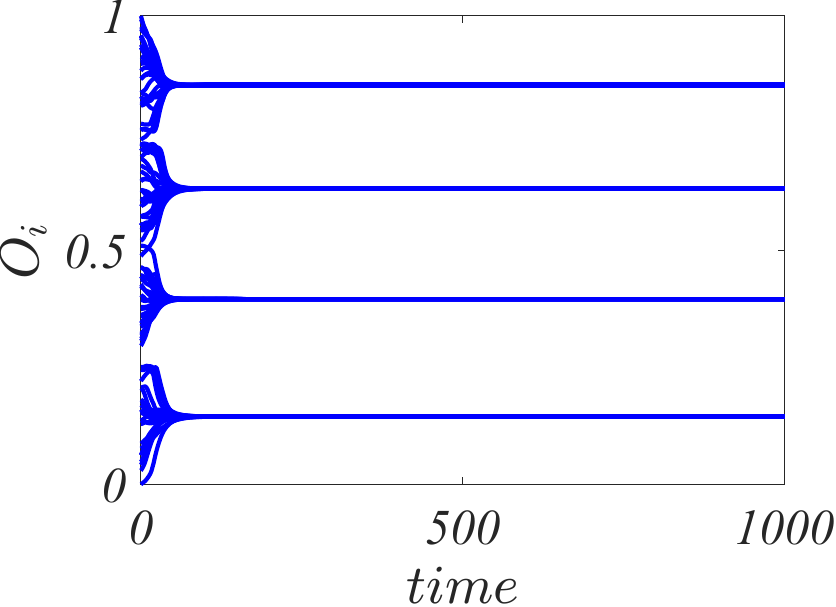} &
        \includegraphics[width=0.19\textwidth]{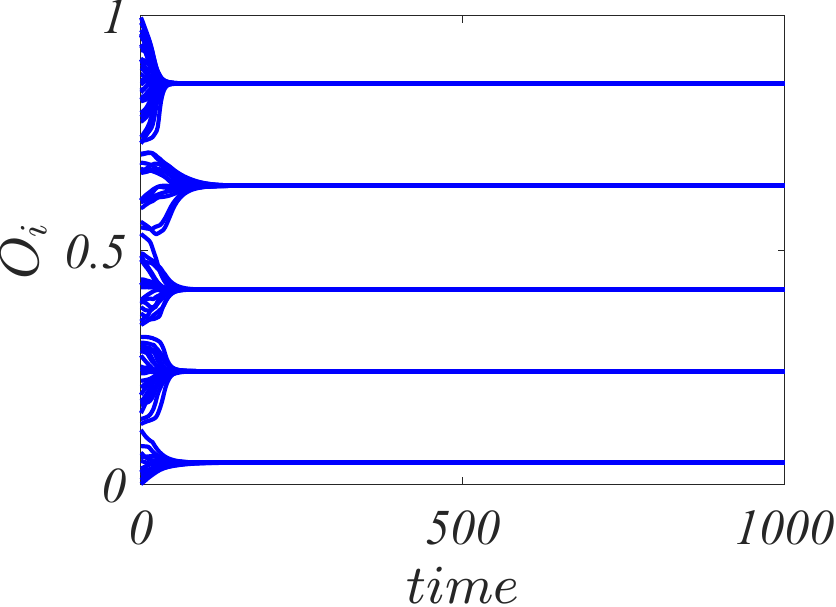}\\
        \textbf{(a2)} & \textbf{(b2)} & \textbf{(c2)} & \textbf{(d2)} & \textbf{(e2)} \\

        \includegraphics[width=0.19\textwidth]{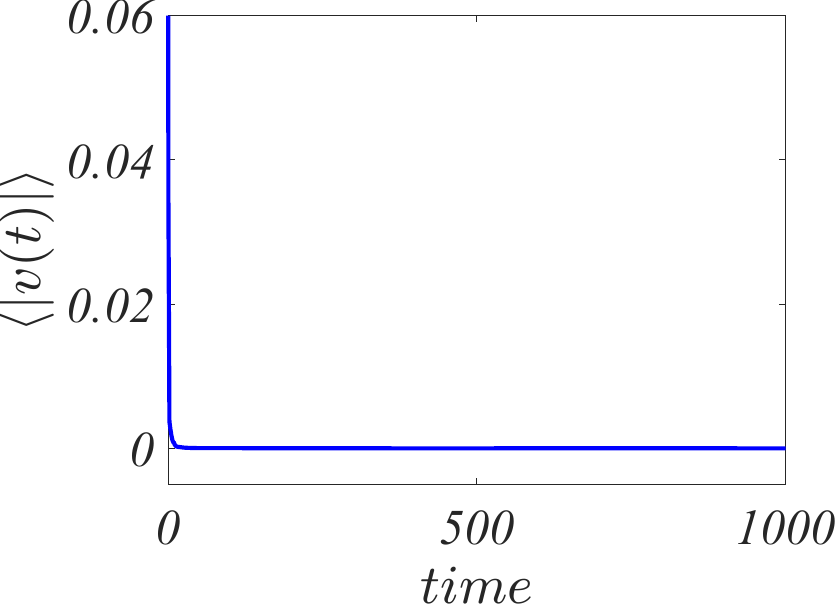} &
        \includegraphics[width=0.19\textwidth]{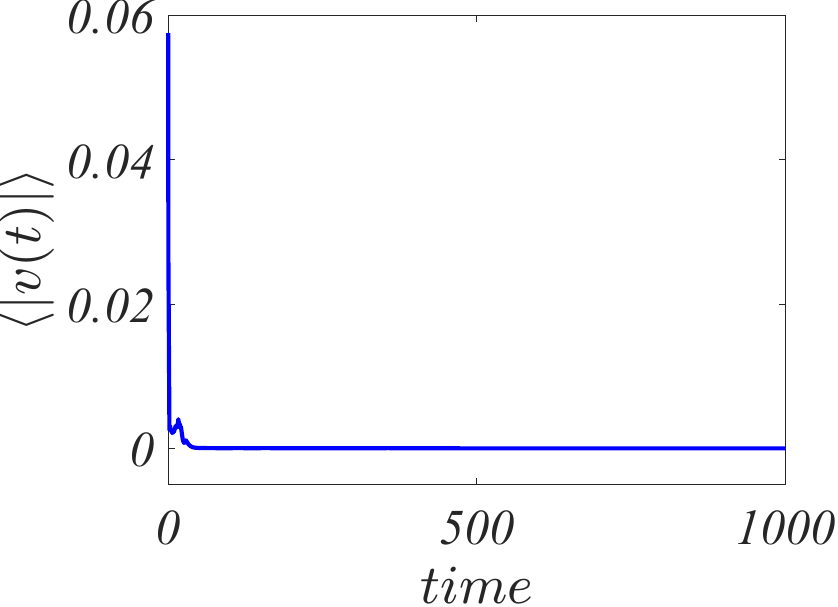} &
        \includegraphics[width=0.19\textwidth]{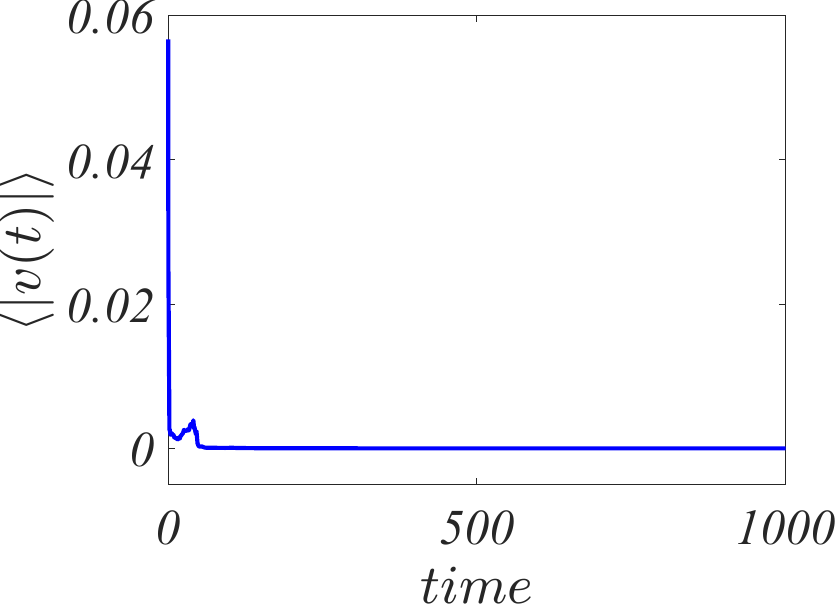} &
        \includegraphics[width=0.19\textwidth]{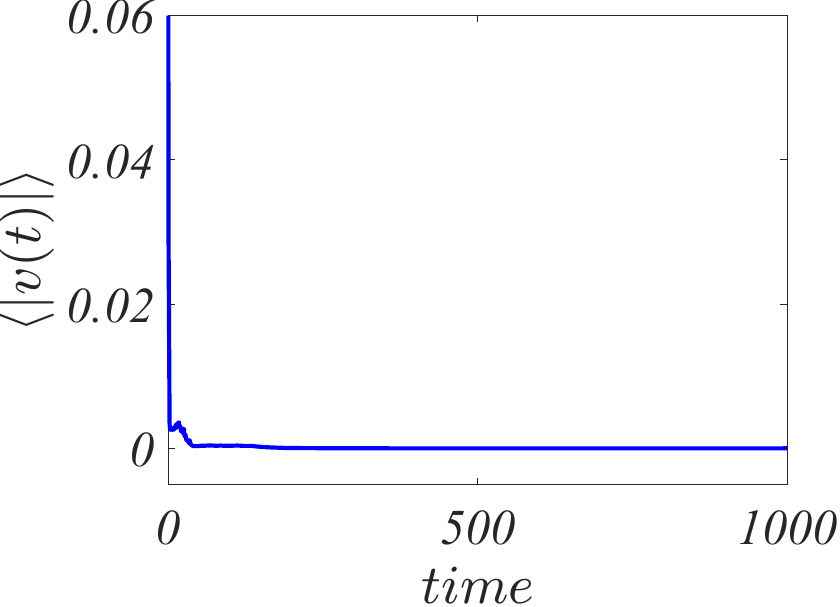} &
        \includegraphics[width=0.19\textwidth]{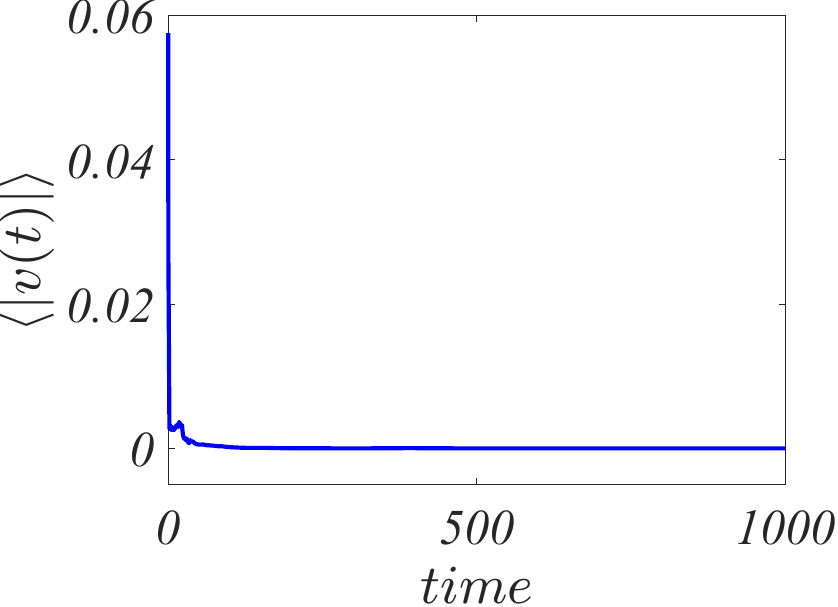}\\
        \textbf{(a3)} & \textbf{(b3)} & \textbf{(c3)} & \textbf{(d3)} & \textbf{(e3)}
    \end{tabular}
    \caption{{\bf Effect of the bounded-confidence threshold $d$ for the linear attraction kernel}. From left to right, $d=\frac{1}{2}$, $\frac{1}{4}$, $\frac{1}{6}$, $\frac{1}{8}$, and $\frac{1}{10}$. The first row shows the final spatial configuration, the second row the opinion dynamics, and the third row the average swarm velocity. The remaining parameters are $J=1$, $N=100$, $\mu=0.2$, and $\lambda=3$.}
    \label{fig:fig1_LK}
\end{figure*}

\begin{figure*}[ht!]
    \centering
    \begin{tabular}{cccc}
        \includegraphics[width=0.23\textwidth]{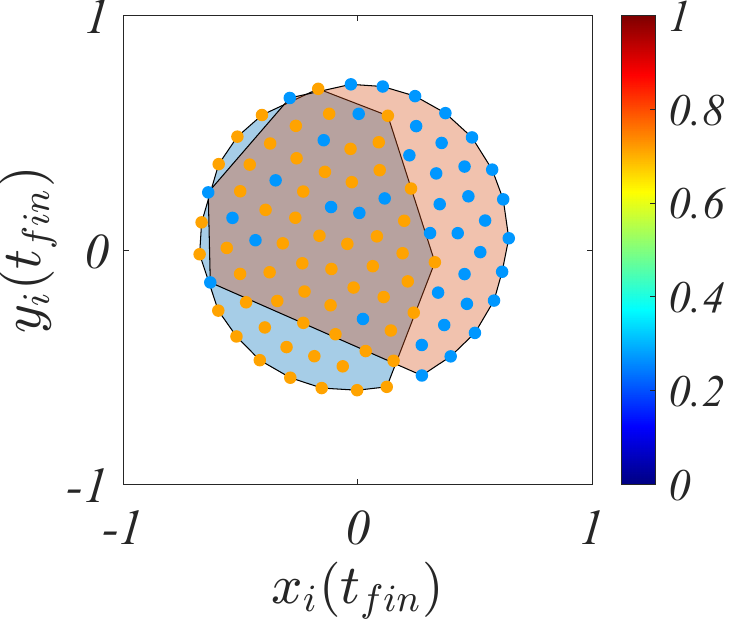} &
        \includegraphics[width=0.23\textwidth]{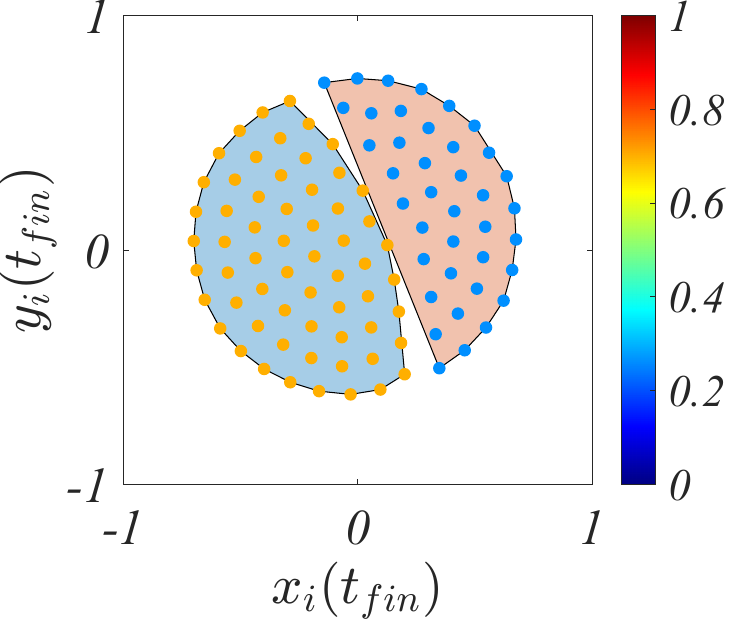} &
        \includegraphics[width=0.23\textwidth]{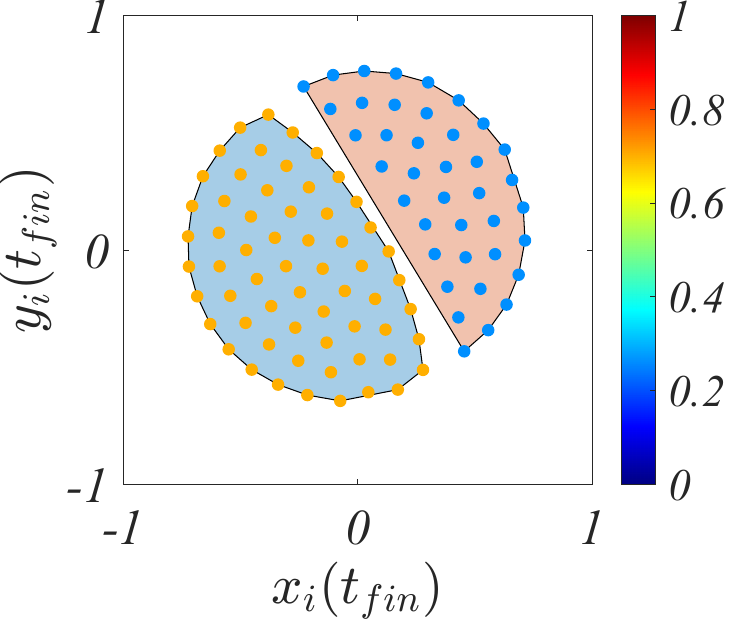} &
        \includegraphics[width=0.23\textwidth]{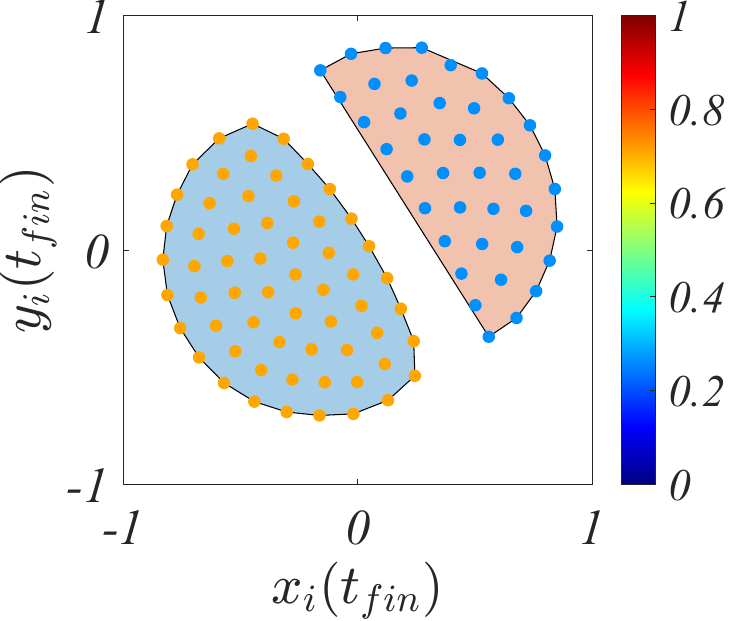}\\
        \textbf{(a1)} & \textbf{(b1)} & \textbf{(c1)} & \textbf{(d1)}\\

        \includegraphics[width=0.23\textwidth]{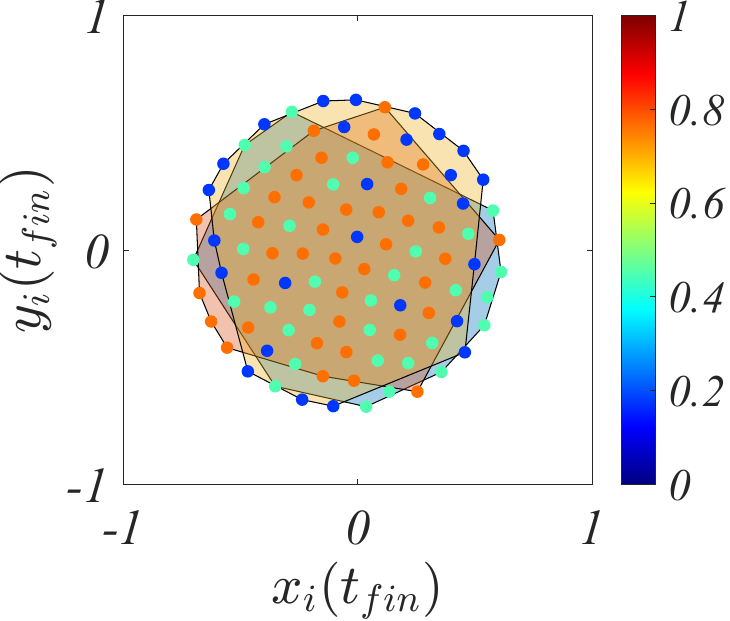} &
        \includegraphics[width=0.23\textwidth]{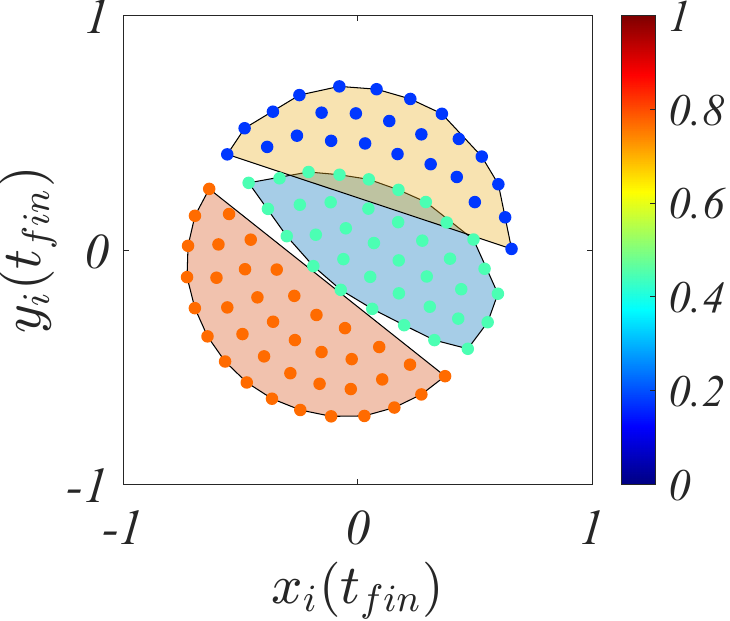} &
        \includegraphics[width=0.23\textwidth]{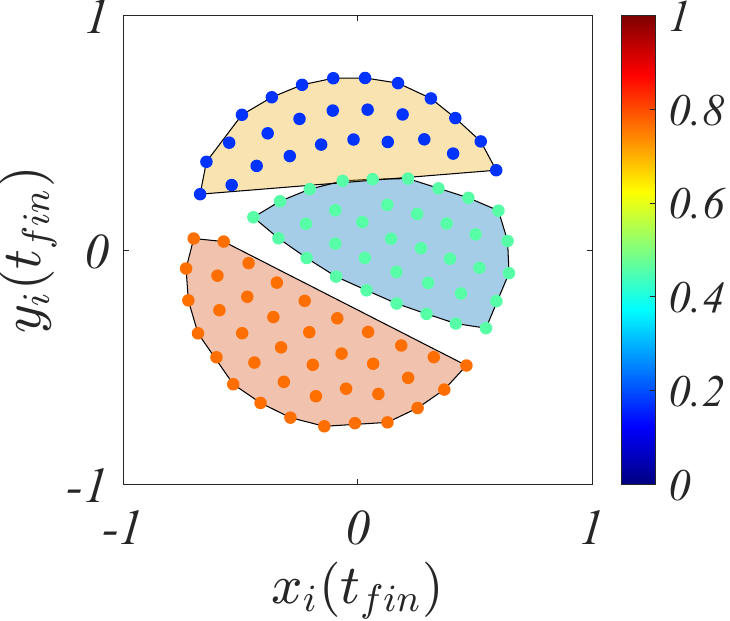} &
        \includegraphics[width=0.23\textwidth]{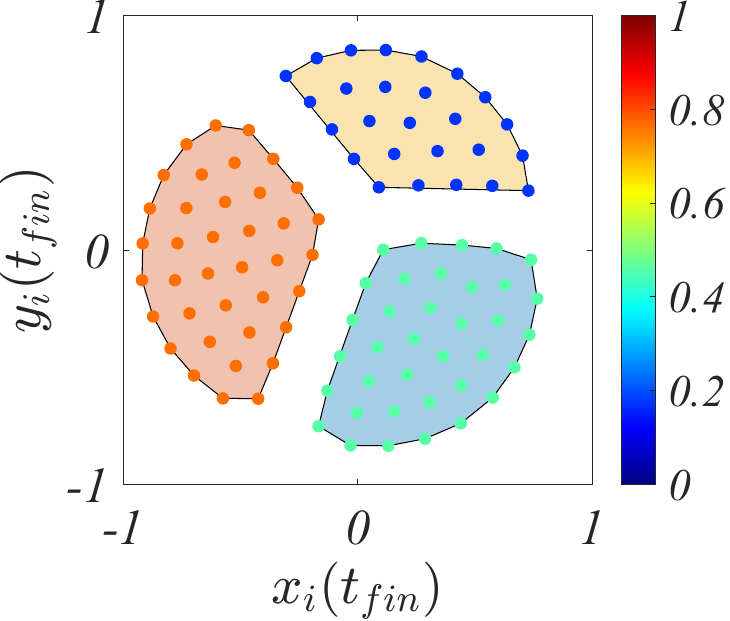}\\
        \textbf{(a2)} & \textbf{(b2)} & \textbf{(c2)} & \textbf{(d2)}\\

        \includegraphics[width=0.23\textwidth]{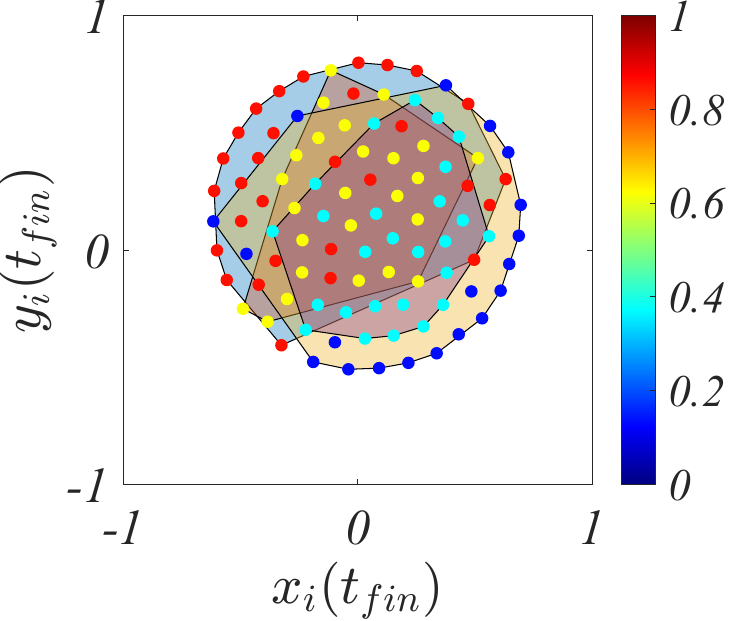} &
        \includegraphics[width=0.23\textwidth]{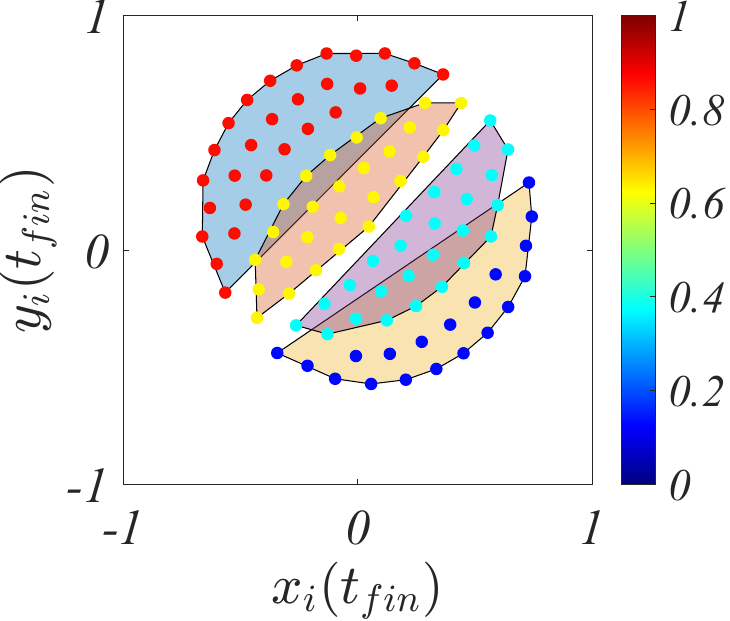} &
        \includegraphics[width=0.23\textwidth]{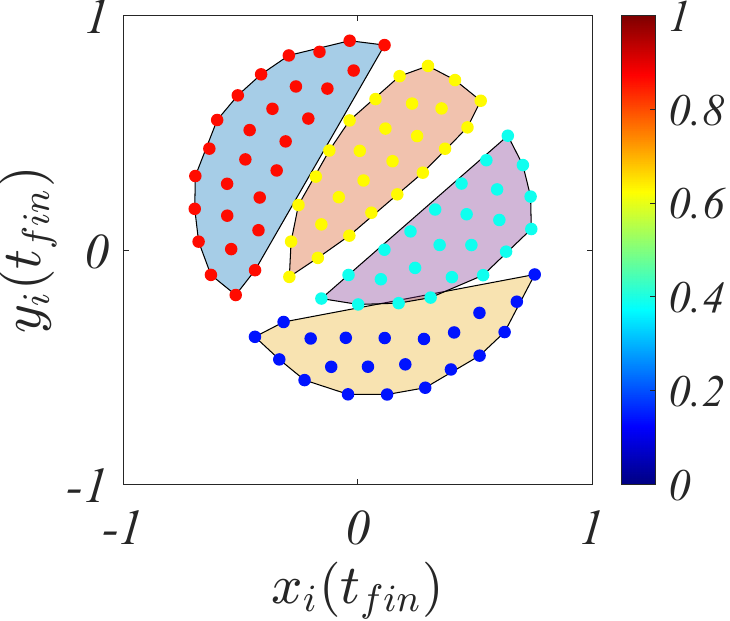} &
        \includegraphics[width=0.23\textwidth]{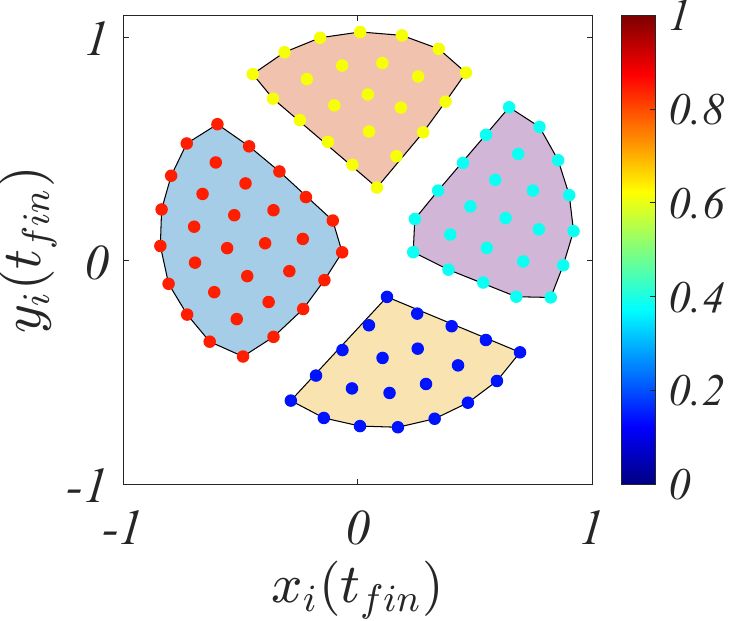}\\
        \textbf{(a3)} & \textbf{(b3)} & \textbf{(c3)} & \textbf{(d3)}\\

        \includegraphics[width=0.23\textwidth]{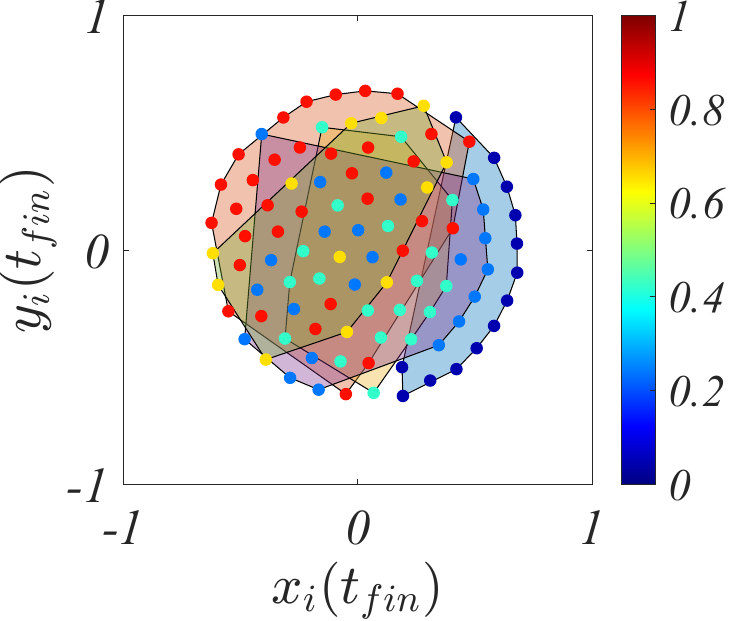} &
        \includegraphics[width=0.23\textwidth]{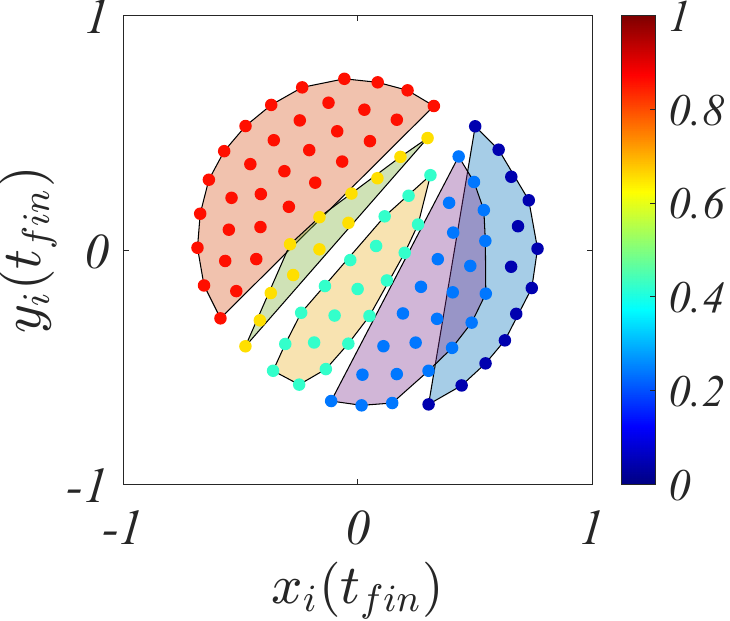} &
        \includegraphics[width=0.23\textwidth]{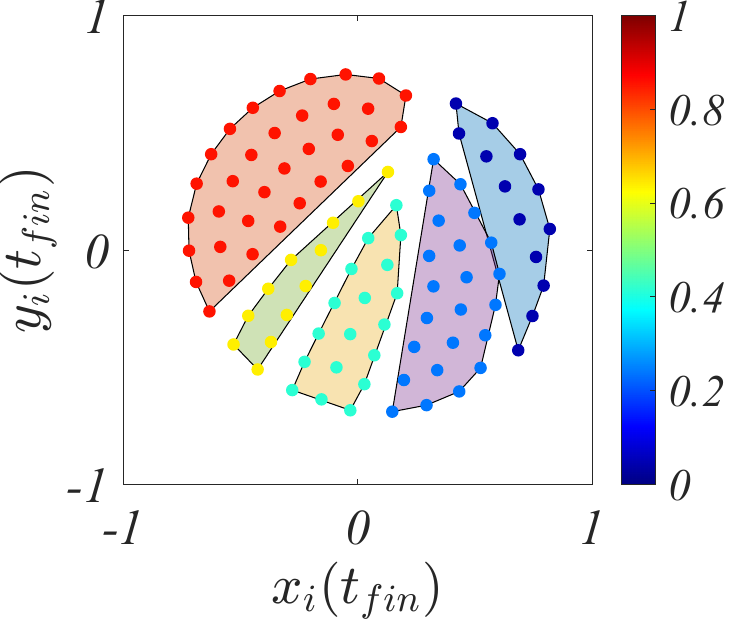} &
        \includegraphics[width=0.23\textwidth]{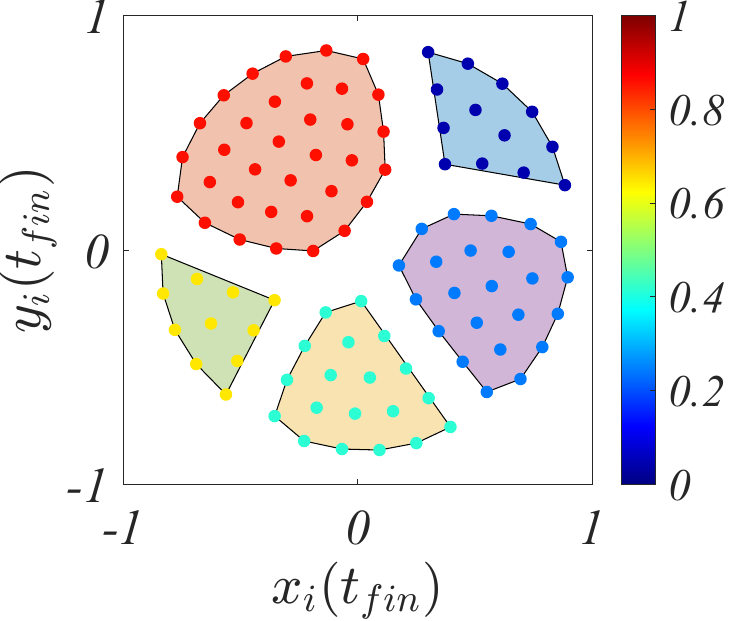}\\
        \textbf{(a4)} & \textbf{(b4)} & \textbf{(c4)} & \textbf{(d4)}
    \end{tabular}
    \caption{Combined effect of $d$ and $\lambda$ for the linear attraction kernel. Rows correspond to $d=\frac{1}{4}$, $\frac{1}{6}$, $\frac{1}{8}$, and $\frac{1}{10}$ from top to bottom; columns correspond to $\lambda=0.1$, $0.8$, $1.5$, and $10$ from left to right. The remaining parameters are $J=1$, $N=100$, and $\mu=0.2$.}
    \label{fig:fig2_LK}
\end{figure*}

In this section we report some numerical results to show the impact of the main model parameters on the dynamical behavior. Figures~\ref{fig:fig1_LK} and~\ref{fig:fig2_LK} display the counterparts of Figures~\ref{fig:fig1} and~\ref{fig:fig2} discussed in the main text, for the present case of linear-kernel. Their aim is to show the robustness of the dynamical behaviors, rather than a separate qualitative discussion. The main observation is unchanged: decreasing $d$ increases the number of opinion groups, while increasing $\lambda$ favors spatial separation between groups with different opinions.

\begin{figure*}[ht!]
    \centering
    \begin{tabular}{cccc}
        \includegraphics[width=0.23\textwidth]{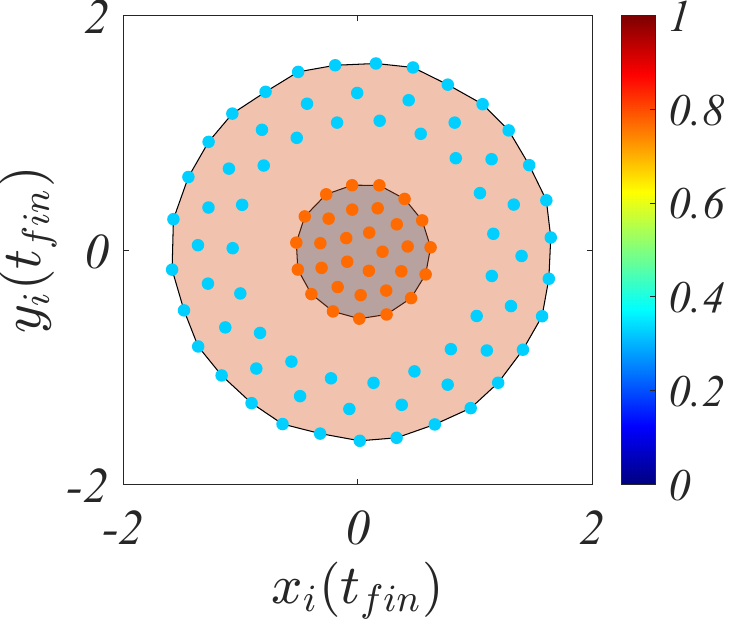} &
        \includegraphics[width=0.23\textwidth]{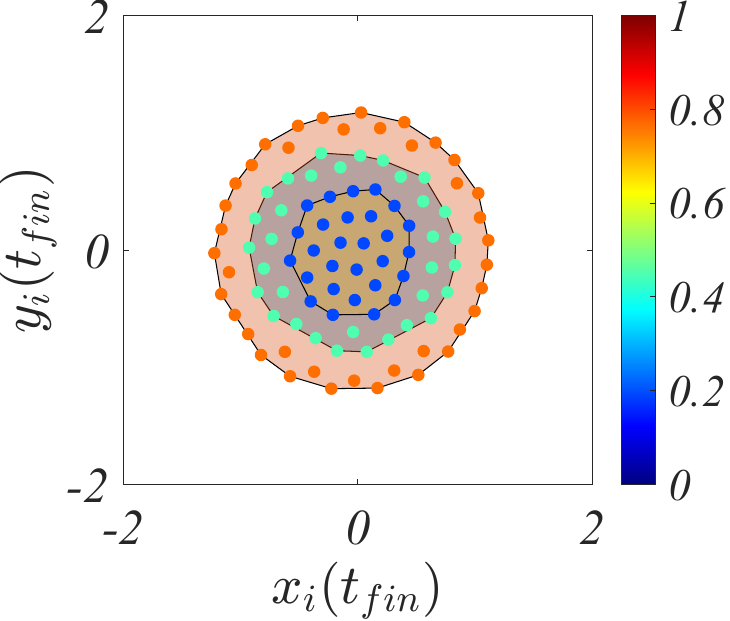} &
        \includegraphics[width=0.23\textwidth]{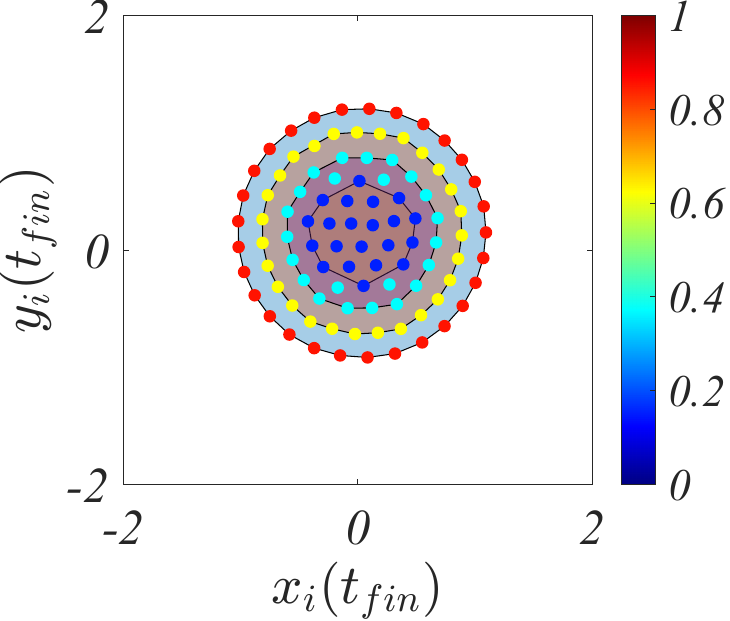} &
        \includegraphics[width=0.23\textwidth]{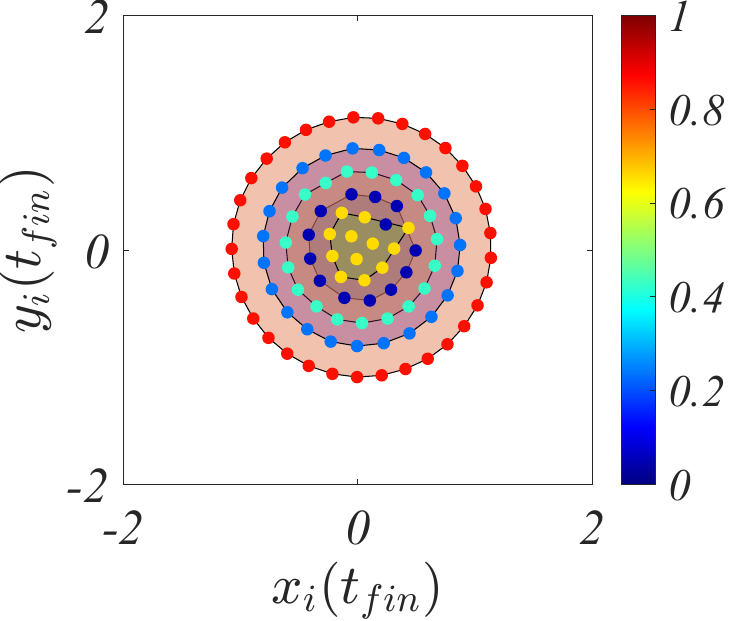}\\
        \textbf{(a)} & \textbf{(b)} & \textbf{(c)} & \textbf{(d)}
    \end{tabular}
    \caption{{\bf Static instructed cluster configurations for the linear kernel with $\lambda=30$}. Panels correspond to: (a) $d=\frac{1}{4}$, $J=-1$; (b) $d=\frac{1}{6}$, $J=-1$; (c) $d=\frac{1}{8}$, $J=-0.8$; and (d) $d=\frac{1}{10}$, $J=-1$. The remaining parameters are $N=100$ and $\mu=0.2$.}
    \label{fig:figSIC}
\end{figure*}

Despite those similarities, the linear-kernel model exhibits a new behavior; for negative values of $J$ close to $-1$ and sufficiently large $\lambda$, the linear kernel can also produce static nested clusters (see Fig. \ref{fig:figSIC}). In this regime, opinion groups are distinct but their spatial supports need not be fully separated, there is a ``central group'', i.e., forming a small disk, surrounded by other groups spatially organized in a ring shape. 

\subsection{Radius of the consensus state}
\label{app:linear_kernel_radius}

\begin{figure*}[ht!]
    \centering
    \includegraphics[width=0.4\textwidth]{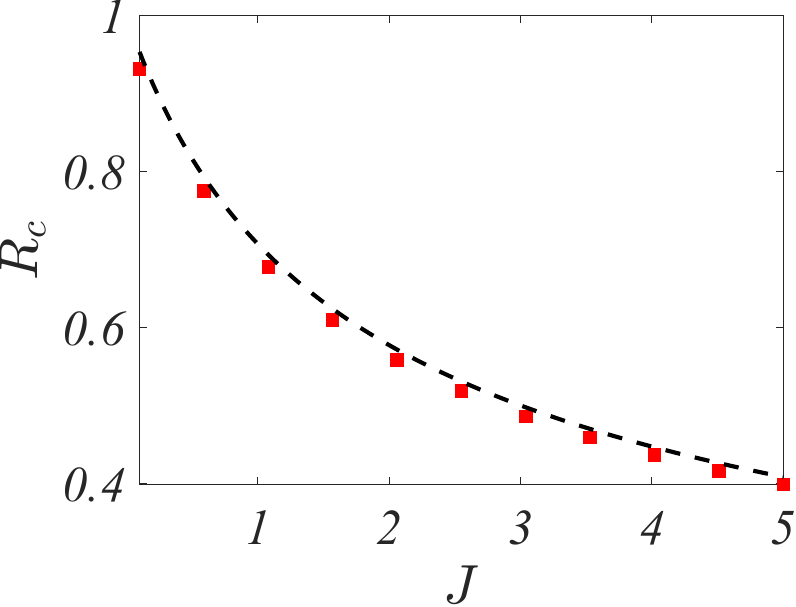}
    \caption{{\bf Radius of the consensus state as a function of $J$ for the linear kernel.} The red squares correspond to simulation results, while the black dashed line corresponds to Eq.~\eqref{eq:radius_sync_lk} with $A=1$ and $B=1$.}
    \label{fig:synch_radius_vs_J}
\end{figure*}

The linear kernel allows a simple estimate of the radius of the fully consensus state, by following ideas developed in~\cite{o2017oscillators}. Assume thus all agents to share the same opinion, so that $e^{-\lambda |O_{ij}|}=1$, and that the stationary spatial distribution is radially symmetric with density $\rho(\mathbf r)$, as confirmed by the numerical simulations. In the continuum limit, i.e., $N\rightarrow \infty$, the velocity field is given by
\begin{equation}
\mathbf v(\mathbf r)=
\int
\left[
(A+J)(\tilde{\mathbf r}-\mathbf r)
-B\frac{\tilde{\mathbf r}-\mathbf r}{\|\tilde{\mathbf r}-\mathbf r\|^2}
\right]\rho(\tilde{\mathbf r})\,d\tilde{\mathbf r}.
\label{eq:linear_kernel_velocity_field}
\end{equation}
Because the equilibrium is stationary, we have $\mathbf v=0$ and hence $\nabla\cdot\mathbf v=0$. By using the formula relating the divergence of $(\mathbf r-\tilde{\mathbf r})/\|\mathbf r-\tilde{\mathbf r}\|^2$ to the $\delta$-function valid in two dimensions, i.e.,
\begin{equation*}
\nabla\cdot\left[\frac{(\mathbf r-\tilde{\mathbf r})}{\|\mathbf r-\tilde{\mathbf r}\|^2}\right]=2\pi\delta(\mathbf r-\tilde{\mathbf r})\, , 
\end{equation*}
into~\eqref{eq:linear_kernel_velocity_field}, one obtains
\begin{equation}
-2(A+J)+2\pi B\rho(\mathbf r)=0\,.
\label{eq:linear_kernel_density_balance}
\end{equation}
Thus the consensus state has constant density
\begin{equation}
\rho_s(\mathbf{r})=\frac{A+J}{\pi B},
\label{eq:linear_kernel_density}
\end{equation}
provided $A+J>0$. Since the density is normalized and zero outside a disk of radius $R_c$, it is straightforward to deduce the value of the latter
\begin{equation}
R_c=\sqrt{\frac{B}{A+J}}\,.
\label{eq:radius_sync_lk}
\end{equation}
In Fig.~\ref{fig:synch_radius_vs_J} we compare the analytical expression for $R_c$ with the one obtained from numerical simulations and we can conclude that the agreement is excellent.

\subsection{Three-dimensional extension}
\label{app:linear_kernel_3D}

\begin{figure*}[ht!]
    \centering
    \begin{tabular}{cc}
        \includegraphics[width=0.35\textwidth]{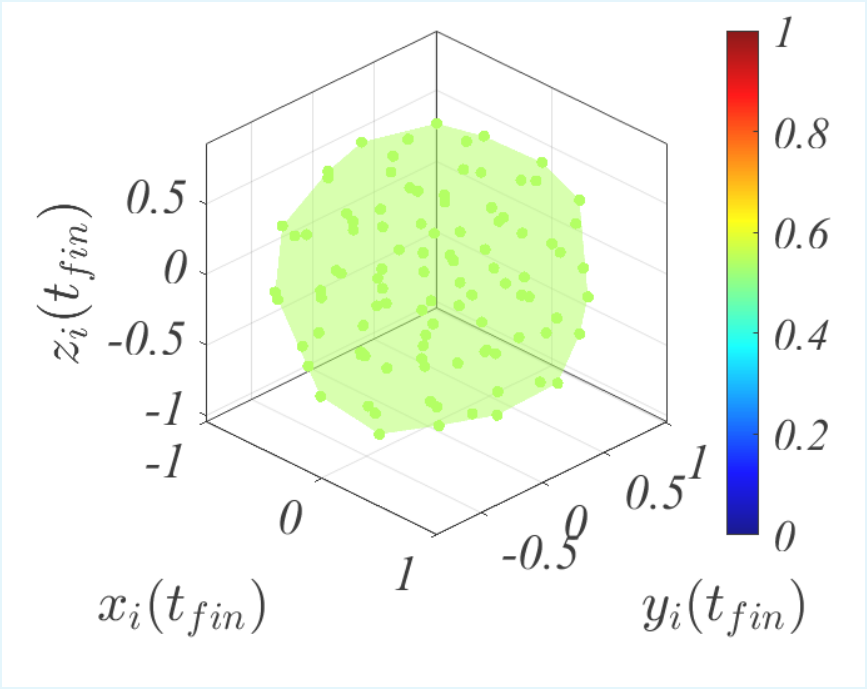} &
        \includegraphics[width=0.35\textwidth]{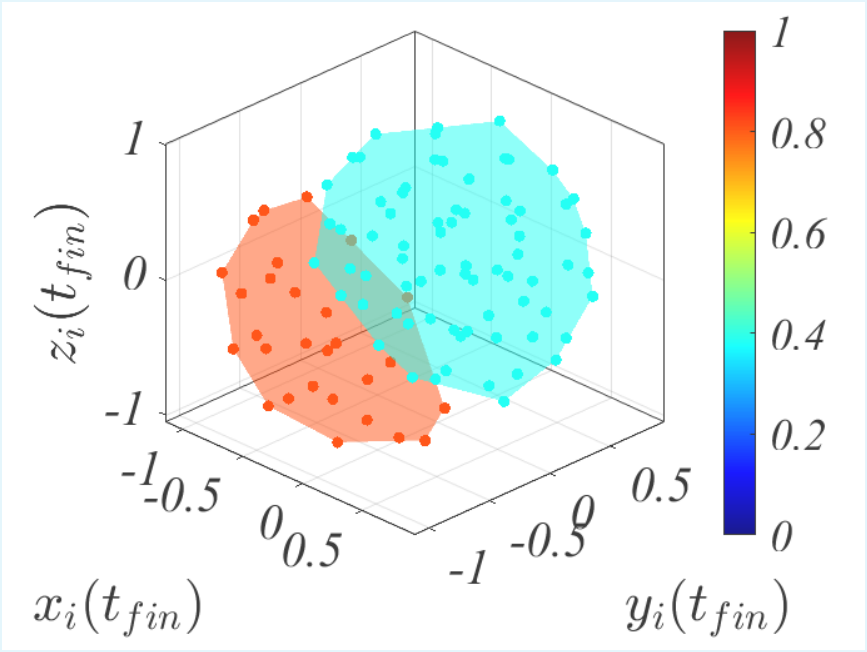}\\
        \textbf{(a)} & \textbf{(b)}\\
        \includegraphics[width=0.35\textwidth]{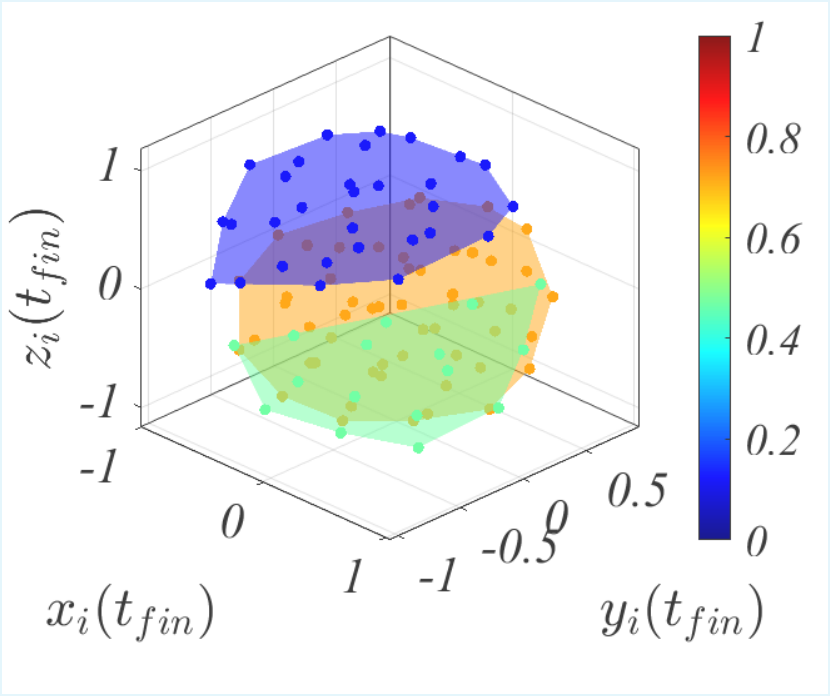} &
        \includegraphics[width=0.35\textwidth]{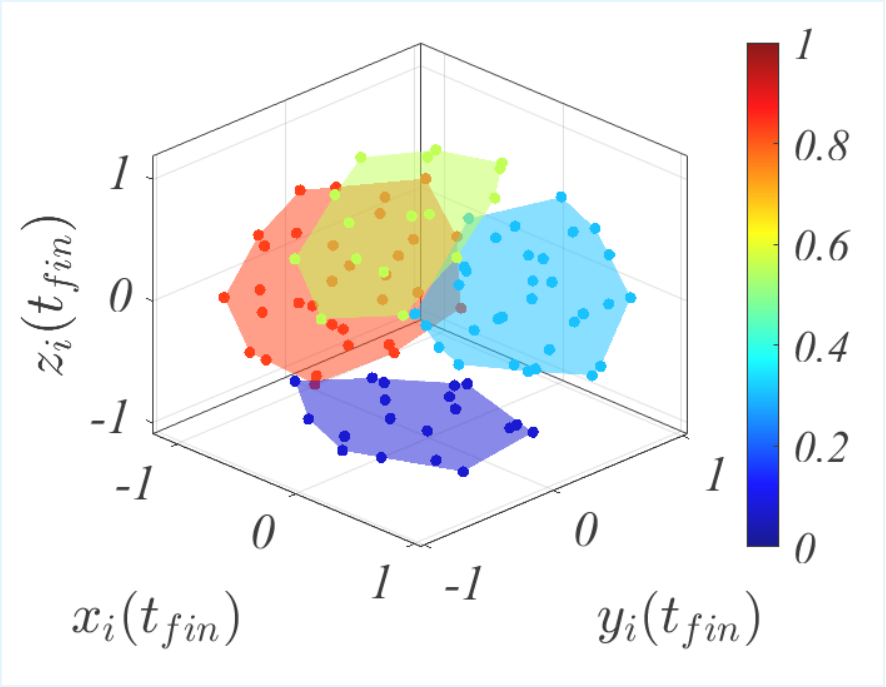}\\
        \textbf{(c)} & \textbf{(d)}
    \end{tabular}
    \caption{{\bf Three-dimensional linear-kernel model}. We show the swarm configurations at time $t_{fin}=500$ where agents are colored according to their opinion. Panels correspond to (a) $d=\frac{1}{2}$, (b) $d=\frac{1}{4}$, (c) $d=\frac{1}{6}$, and (d) $d=\frac{1}{8}$. The remaining parameters are $J=2$, $\lambda=10$, $N=100$, and $\mu=0.2$.}
    \label{fig:swarmalators3D}
\end{figure*}

The aim of this section is to briefly present an extension of the model with linear-kernel to the three dimensional case by defining $\mathbf r_i=(x_i,y_i,z_i)\in\mathbb R^3$. The model is thus giving by 
\begin{align}
\dot{\mathbf r}_i
&=\frac{1}{N}\sum_{j\neq i}
\left[
\mathbf r_{ij}\left(A+J e^{-\lambda |O_{ij}|}\right)
-B\frac{\mathbf r_{ij}}{r_{ij}^{3}}
\right],
\label{eq:linear_kernel_3D_spatial}\\
\dot O_i
&=\frac{\mu}{N}\sum_{j\neq i}
\frac{\Theta\left(d-|O_{ij}|\right)O_{ij}}{r_{ij}}\, .
\label{eq:linear_kernel_3D_opinion}
\end{align}
The numerical results reported in Fig.~\ref{fig:swarmalators3D} show that the dependence on $d$ remains qualitatively the same in three dimensions: smaller confidence thresholds produce more opinion groups and, for the parameters considered here, more spatially separated clusters.

\begin{figure*}[ht!]
    \centering
    \begin{tabular}{cc}
        \includegraphics[width=0.4\textwidth]{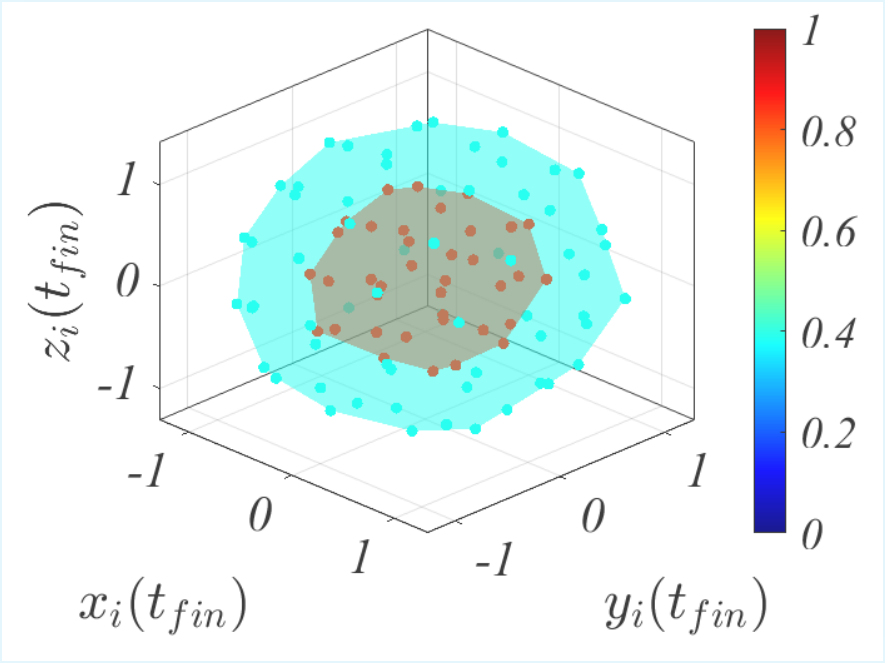} &
        \includegraphics[width=0.4\textwidth]{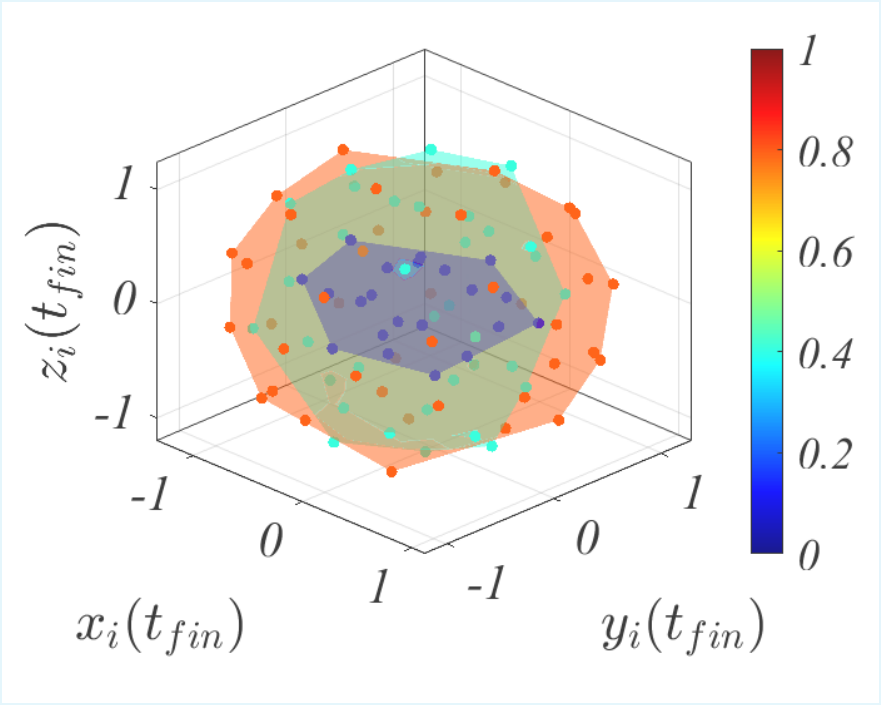}\\
        \textbf{(a)} & \textbf{(b)}
    \end{tabular}
    \caption{{\bf Static nested clusters in three dimensions for $\lambda=30$ and $J=-0.95$}. Panels correspond to (a) $d=\frac{1}{4}$ and (b) $d=\frac{1}{6}$. The remaining parameters are set to $N=100$ and $\mu=0.2$.}
    \label{fig:figSIC3D}
\end{figure*}

Static nested clusters can also be observed in three dimensions, as illustrated in Fig.~\ref{fig:figSIC3D}. These states are the three-dimensional analogue of Fig.~\ref{fig:figSIC}: the opinion groups are distinguishable, but their spatial organization is not simply a collection of disconnected components, they assume an ``onion''--like structure, generalizing thus the ring in two dimension.

\end{document}